\documentclass[12pt]{article}

\usepackage{tikz}
\usetikzlibrary{quantikz}
\usepackage{ifxetex,ifluatex,fullpage}
\usepackage{hyperref,graphicx,float,amsmath,amsfonts,amssymb,braket}
\if\ifxetex T\else\ifluatex T\else F\fi\fi T%
  \usepackage{fontspec}
\else
  \usepackage[T1]{fontenc}
  \usepackage[utf8]{inputenc}
  \usepackage{lmodern}
  \usepackage{url}
\fi
\usepackage{authblk}
\usepackage{hyperref}
\usepackage{cite}

\title{Error Resilient Quantum Amplitude Estimation from Parallel Quantum Phase Estimation}
\author[1]{M.C. Braun$^*$}
\author[1]{T. Decker\footnote{For contact email: thomas.decker@jos-quantum.de or sven.kerstan@jos-quantum.de. Authors are listed in alphabetical order.}}
\author[1]{N. Hegemann$^*$}
\author[1]{S.F. Kerstan$^*$}
\affil[1]{JoS QUANTUM Intellectual Property GmbH\\Frankfurt am Main\\Germany}
\begin{document}
\maketitle
%\tableofcontents

We show how phase and amplitude estimation algorithms can be parallelized. This can reduce the gate depth of the quantum circuits to that of a single phase kickback of an operator with a small overhead.
Further, we show that for quantum amplitude estimation, the parallelization can lead to improvements in resilience against quantum errors. The resilience is not caused by the lower gate depth, but by the structure of the algorithm. The parallelization and the results on error resilience hold for the standard version and for low depth versions of phase estimation and quantum amplitude estimation. Methods presented are subject of a patent application [Quantum computing device: Patent application EP 21207022.1].\\

\section{Introduction}
\begin{figure}[H] \label{kickbackspreview}
\centering

\includegraphics[width=0.8\textwidth]{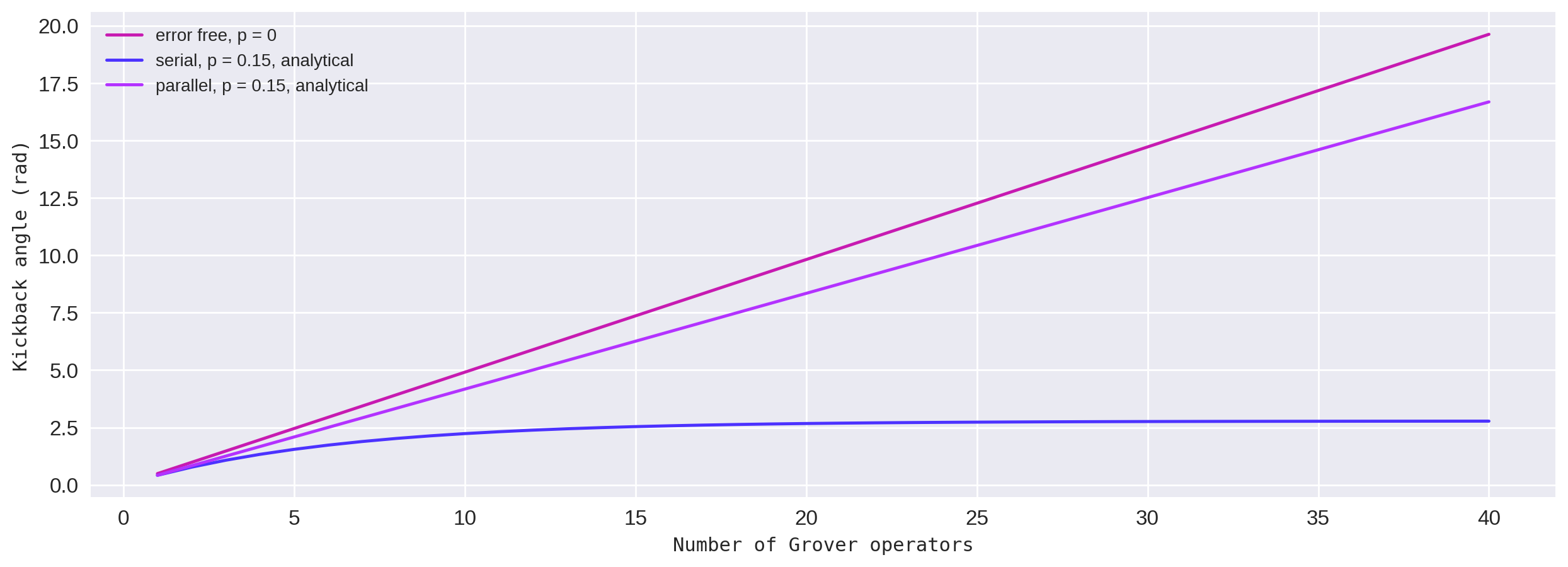}
\caption{The expected kickback angle generated by a collection of Grover operators in the error free case, in the noisy case with serial execution of the Grover operators and in the noisy case with parallel execution, respectively. Each Grover operator picks up an error with probability $p$.}
\end{figure}

Many potential commercial applications for quantum computers rely on quantum algorithms that can execute the quantum equivalent of Monte-Carlo simulations quadratically faster than classical Monte-Carlo implementations on classical computers. Well-known examples are Value-at-Risk (VaR) calculations \cite{quantumriskanalysis}, the calculation of option prices \cite{quantumoptionpricing} or the calculation of sensitivities of risk models \cite{jos}, all of which are based on the Quantum Amplitude Estimation algorithm (QAE, see~\cite{QAE}), which itself is a variant of the Quantum Phase Estimation algorithm (QPE, see~\cite{shor}).\\
Unfortunately, quantum computers, which satisfy the fidelity requirements for such applications~\footnote{Practical applications with fidelities below 99.99999\% seem unrealistic as calculations based on \cite{jos} and \cite{quantumadvoptionpric} indicate.}, are not available today and are not expected to be available in the foreseeable future (see \cite{ibmroadmap, googleroadmap, rigettiroadmap, ionqroadmap}). Since the number of noisy qubits is expected to grow quickly compared to the fidelity (see \cite{ibmroadmap, googleroadmap, rigettiroadmap, ionqroadmap}), it may be worthwhile to explore new ways to use extra qubits to make known quantum algorithms more error resilient. Extra qubits can be used for standard, universal error correction methods \cite{NC}, which result in error resilience, but we might hope that in some cases, properties particular to a certain kind of quantum circuit might be exploited to achieve error resilience with less overhead than universal error correction.\\
In this context, we present parallel versions of QAE, which are based on parallel versions of QPE. Apart from the advantages that come with the vastly reduced circuit depth (see section \ref{secIndParallel}), i.e. a reduced execution time and therefore also fewer decoherence issues, we show that the effect of unitary errors on the register on which the Grover operator works is reduced compared to the standard, serial execution of the Grover operators in QAE and low depth variants of it. To illustrate this, an instructive comparison is depicted in figure~\ref{kickbackspreview} for the type of operators that we consider in section~\ref{secErrorGrover}. The figure shows how the kickbacks to a single qubit from Grover operators, each of which picks up an error with probability $p$ (right before or after the Grover operator), accumulate\footnote{This is the basic building block of of circuits for both, QAE and low depth versions of QAE.}.
In the error free case, the relation between the total kickback angle and the number of Grover operators is linear with slope $2\theta$. In section \ref{errorsinamplitudeestimation} we show that in the presence of errors, which have a probability of $p$, in the Grover register (but not inside the Grover operators), the number of successful kickbacks is drawn from a geometric distribution and the expected value for the sum of the kickbacks cannot exceed $2\theta \frac{1-p}{p}$,
regardless of how many Grover operators we execute. This sets a bound on the accuracy of serial QAE (standard or low depth), which cannot be exceeded: The number of effective Grover operators is always smaller than
\begin{equation} \label{Nbound}
N_{\rm eff}=\frac{1-p}{p}\,.
\end{equation}
This cannot be improved by an increase of the number of times the circuit is measured. Consider an example: If there is a 20\% probability for an error, then the precision of any form of serial QAE will be limited to a single decimal digit\footnote{The resolution of the result of QAE is determined by the highest number of Grover oracles kicking back to a single qubit. According to equation \ref{Nbound}, a 20\% error corresponds to 4 effective Grover oracles. The third binary digit of the QAE result can be determined with kickbacks of up to 4 Grover operators, and has a decimal value of 0.125. So the maximal resolution is lower than 0.1, i.e. a single decimal digit.}, regardless of how many times we measure or how we post-processes the results of measurements. This precision is several orders of magnitude away from what is required for typical practical applications.\\
In contrast to this, we show in section \ref{errorsinldqae} that the parallel version still exhibits the linear growth and the equivalent to equation \ref{Nbound} is
\begin{equation} \label{Nbound2}
N_{\rm eff}=(1-p) N\,.
\end{equation}
This means that after estimating the error probability $p$, we can in principle achieve any desired precision\footnote{The number of shots will generally have to increase for higher precision and, depending on the example, this can diminish the quantum advantage, and in some cases destroy it.}, a big advantage over the standard algorithms.\\
This paper is organized as follows: In section~\ref{sectionStandardPhaseEstimation}, we recap standard QPE and QAE and analyze a key issue: the effect of errors in Grover operators and on the register on which the operators act.
In section~\ref{sectionParallelPhaseEstimation}, we introduce three different ways in which a parallelization of circuits for QPE can be implemented: the simple version, which does not fully minimize the gate depth of the circuits, the indirect version, which is based on entanglement and has the minimal gate depth, and the implementation by reinitializing qubits, which is useful if the number of available independent qubits is too low for the first two versions.
In section~\ref{sectionParalleQAE} we show how the preceding sections apply to QAE and low depth versions of QAE, including the behavior when errors occur.
In section~\ref{sectionEP} we show how single eigenstates can be prepared for QAE in some cases.
In section~\ref{sectionErrorCorrection} we introduce a technique for QAE specific error correction.
In section~\ref{sectionExample}, we apply the techniques -- the eigenstate preparation, the parallelization and the QAE specific error correction -- to a small scale version of a real life business problem.
The appendix contains results for the special case of Grover operators on a single qubit, which is different from the general case.\\
After the first version of this manuscript had been completed, it has come to our attention that Knill, Ortiz and Somma had presented one of our parallel implementations, the entangled version of the standard QAE circuit, in \cite{Knill}. In that work, they also presented an approach for the eigenstate preparation, which complements what we describe in section~\ref{sectionEP}.

\section{Standard Phase and Amplitude Estimation}\label{sectionStandardPhaseEstimation}
Suppose that we have a quantum circuit implementation for some operator $G$ and also one for an operator EP, which produces an eigenstate of $G$. Then there is a quantum algorithm that can calculate the eigenvalue of the eigenstate prepared by EP efficiently: the quantum phase estimation algorithm.\\
The phase estimation algorithm is well-known and is described in standard textbooks, such as~\cite{NC}. For example, a circuit that calculates the eigenvalue of the operator $G$ that acts on a quantum register consisting of several qubits with three bits precision on a quantum system looks like this:
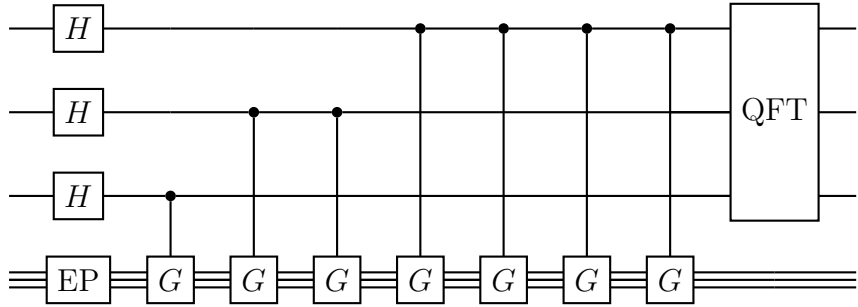
\begin{figure}[H]
\centering
\begin{quantikz}
\qw&\gate{H}              &\qw              &\qw              &\qw              &\ctrl{3}         &\ctrl{3}         &\ctrl{3}         &\ctrl{3}         &\gate[wires=3]{\rm QFT}&\qw\\
\qw&\gate{H}              &\qw              &\ctrl{2}         &\ctrl{2}         &\qw              &\qw              &\qw              &\qw              &\qw                    &\qw\\
\qw&\gate{H}              &\ctrl{1}         &\qw              &\qw              &\qw              &\qw              &\qw              &\qw              &\qw                    &\qw\\
\qwbundle[alternate]{}&\gate{\rm EP}\qwbundle[alternate]{}&\gate{G}\qwbundle[alternate]{}&\gate{G}\qwbundle[alternate]{}&\gate{G}\qwbundle[alternate]{}&\gate{G}\qwbundle[alternate]{}&\gate{G}\qwbundle[alternate]{}&\gate{G}\qwbundle[alternate]{}&\gate{G}\qwbundle[alternate]{}&\qwbundle[alternate]{}                    &\qwbundle[alternate]{}\\
\end{quantikz}
\caption{An example for a circuit implementing the standard quantum phase estimation for a unitary operator $G$ on a quantum system, where ${\rm EP}$ prepares an eigenstate $|\lambda\rangle$ of $G$. The eigenvalue $\lambda$ is unknown and the circuit produces an estimation of $\lambda$ with three bit precision.}
\label{PE}
\end{figure}

The gate ${\rm EP}$ prepares an eigenstate of the gate $G$ or a suitable superposition of eigenstates. The output of
the Quantum Fourier Transform (QFT) can then either be measured if the phase estimation is the final result or it can be fed into another quantum circuit for further processing, see e.g. ref.~\cite{jos}, where the result of a QAE is the input of a Grover search.\\
The same circuit performs a QAE if the operator $G$ is a Grover operator (see~\ref{Groveroperators}). This special case is discussed in more detail in the following section.

\subsection{Grover Operators}\label{Groveroperators}
A typical operator $G$ that would occur in relevant applications of QPE is a Grover operator, which takes the form
$G = - M S_0 M^\dagger S_{x}$, e.g. see ref.~\cite{QAE}. Here, $M$ is a unitary operator with $M|0\rangle=|\Psi_0\rangle+|\Psi_1\rangle$ where $|\Psi_0\rangle$ and $|\Psi_1\rangle$ are superpositions of the bad and good states of the computational basis, respectively. We want to find an approximation of the probability $a=\langle \Psi_1|\Psi_1\rangle$ of the good states to occur. The unitary $S_x$ marks the good states with the phase $-1$ and $S_0$ marks the state $|0\rangle$ with the phase $-1$.\\
The structure of the Grover operator leads to special properties~\cite{QAE}: The unitary operator $M$ and its inverse are grouped around the reflection $S_0$ and we can view this as a basis transformation of $S_0$:
\begin{equation}\label{groveroperator}
-M S_0 M^\dagger S_x = - \tilde{S}_0 S_x \;\; {\rm with} \;\; \tilde{S}_0=M S_0 M^\dagger.
\end{equation}
Since two reflections form a rotation, this shows that the Grover operator is a rotation in a 2-dimensional subspace that we call the Grover plane. The eigenvalues of the rotation operation can be written in the form $e^{+i 2\theta}$ and $e^{-i 2\theta}$. The angle $\theta$ is directly related to the quantity we want to measure, i.e. the probability $a$ of the good states:
\begin{equation}\label{eqSinus}
{\rm sin}^2(\theta) = a .
\end{equation}

This structure of the eigenvalues of the Grover operator is the key to the QAE method: If the Grover operator $G$ is performed as a controlled operation, the kickback is $e^{+i 2\theta}$or $e^{-i 2\theta}$, as long as the initial state was prepared in the corresponding eigenstate of $G$.  In the standard version of QAE, it is easy to get around the need to prepare one of the eigenstates, since simply applying $M$ to the zero state $|0\rangle$ produces a superposition of the two non-trivial eigenstates \cite{QAE}. When measuring the output, the superposition collapses to one of the eigenstates, but it does not matter which of the two eigenvectors was realized, since the sign disappears in equation~\ref{eqSinus} by squaring the sine value.\\

Another property of Grover operators that we need to understand for our purposes is the structure of the eigenvalues for the eigenstates that are orthogonal to the Grover plane. It is well-known that these eigenstates all have the eigenvalues $+1$ or $-1$ (see e.g. \cite{QAE}). We need to know how many of $+1$ and $-1$ appear.

If the register of the Grover operator $G$ has $q$ qubits, then the state space for our search has $2^q$ dimensions. The eigenstates form an orthonormal basis of this $2^q$-dimensional space. The two eigenstates with the $\theta$-related eigenvalues form a basis of the Grover plane, and the $2^q-2$ remaining states form a basis of the space orthogonal to the Grover plane.
Since $MS_0 M^\dagger$ acts as the identity on the space orthogonal to the Grover plane, the eigenvalues of $G$ on the orthogonal space are the eigenvalues of $-S_x$, that is $1$ for a good state and $-1$ for a bad state. Let $n_g$ be the number of good states and $n_b$ the number of bad states. If $n_+$ and $n_-$ are the number of eigenvalues $+1$ and $-1$ for eigenstates orthogonal to the Grover plane then this immediately gives us
\begin{eqnarray}\label{numberofeigenvalues}
2^q &=& 2 + n_g + n_b\\ 
n_+ &=& n_g -1\\
n_- &=& n_b -1
\end{eqnarray}
For example, this means that if we have exactly one good state, then we have only eigenstates with $-1$ eigenvalues orthogonal to the Grover plane.

\subsection{Errors and Grover Operators}\label{secErrorGrover}
A key aspect of our work is the discussion of the consequences of errors when QAE is performed. In real world applications like Monte-Carlo simulations, which are too calculation intensive for classical computers, we have Grover oracles on many qubits whereas the number of kickback qubits is much smaller and limited by the requirements of the precision of the estimation. If we consider single qubit errors that are distributed randomly among the qubits then the effect of errors occurring on the qubits on which the Grover operators act is our main concern. As discussed in the previous section, QAE relies on the fact that the quantum register on which a Grover operator acts is in a particular eigenstate which lies in the Grover plane. Some typical hardware errors, such as bit flips ($X$-errors), will generally transform this eigenstate of the Grover operator into another eigenstate or a superposition of eigenvectors that are orthogonal to the Grover plane. We assume that this will be the case here except for Grover operators on one qubit.\\
As stated above, the eigenvalues of a Grover operator are $e^{+2i\theta}$ and $e^{-2i\theta}$ for the eigenstates that span the Grover plane and all other eigenvalues are $+1$ or $-1$ for the eigenstates that are orthogonal to the Grover plane.
A key observation is that the number of eigenvalues $+1$ is equal to the number of good states minus 1, and the number of eigenvalues $-1$ is equal to the number of bad states minus 1, see equation~\ref{numberofeigenvalues}.\\
Let us assume that the probability $a$ of finding a good state is either close to $0$ or close to $1$.
Then the effect of errors will be small in the sense that a randomly occurring error on the Grover register almost certainly causes only a small deviation from the correct kick-back angle $2 \theta$ if one of the two following conditions holds:
\begin{enumerate}
\item{The probability $a$ of finding a good state is very small and the number $n_g$ of good states is much larger than the number $n_b$ of bad states: $n_g \gg n_b$.}
\item{The probability $a$ of finding a good state is very large and the number $n_g$ of good states is much smaller than the number $n_b$ of bad states, $n_g \ll n_b$.}
\end{enumerate}
When the first condition holds, then the correct eigenvalue for the kickback is $e^{2i\theta}=1+\epsilon$ for some small complex $\epsilon$, while in the case of an error the eigenvalue $1$ is kicked back with probability 
$$
\frac{n_+}{n_++n_-} = \frac{n_g-1}{n_g+n_b-2} \approx 1 .
$$
This means that with probability close to 1 the error generates an error in the kickback of just $\epsilon$.\\

In the second case, the correct eigenvalue for the kickback is $e^{2i\theta}=-1+\epsilon$, for some small complex $\epsilon$, while in the case of an error, the eigenvalue $-1$ is kicked back with probability 
$$
\frac{n_-}{n_++n_-} = \frac{n_b-1}{n_g+n_b-2} \approx 1 .
$$
Again, the error this causes in the kickback will be small, that is of size $\epsilon$, with a probability close to 1.\\

Let us look at what happens when the conditions 1 and 2 from above are not met, because the number of good or bad states violates the conditions. We visualize the sum of the kickbacks on the unit circle. Initially, the state sits at $0$ radians. Each correct kickback increases the angle by a small amount of $2\theta$. So after $k$ kickbacks, the state sits at angle $2k \theta$ on the unit circle, still close to the origin if $k$ is sufficiently small. A single error with a $-1$ kickback will add an angle $\pi$ and so the state will end up on the opposite side of the unit circle. At the same time, the effect of any even number $2 s$ of such errors corresponds to just a small error (consisting of not moving the angle further by $4 s \theta$). So with such kickbacks of errors, we lose all control over the magnitude of the total error. The behavior of QAE with errors on the Grover register is discussed in more detail in section~\ref{errorsinamplitudeestimation}. \\
In situations where the conditions 1 or 2 above do not apply, we may still be able to eliminate the big phase errors mentioned above. In particular, in section~\ref{sectionErrorCorrection} we show how to detect and correct such errors for those cases, where the most likely consequence of an error is the kickback of the phase $-1$.

\subsection{Errors and Grover Operators in 2 Dimensions}\label{errorsin2d}
There is a special case of Grover operators which induces a qualitatively different behavior in amplitude estimations: Grover operators in 2 dimensions. First of all, in 2 dimensions the Grover plane is the entire search space, so there are no eigenvalues $+1$ and $-1$ at all. This means that an error cannot move the state of the Grover register out of the Grover plane.
In particular, consider an operator 
$$
G=\left(\begin{array}{cc}{\rm cos}(\alpha) & -{\rm sin}(\alpha) \\
{\rm sin}(\alpha) & {\rm cos}(\alpha) \end{array}\right)
$$ 
which is a rotation in 2 dimensions. For the eigenvectors we then have 
$$
|\Psi_+\rangle=\frac{1}{\sqrt{2}}\left(\begin{array}{c}1\\i\end{array}\right) 
\quad {\rm and} \quad |\Psi_-\rangle=\frac{1}{\sqrt{2}}\left(\begin{array}{c}i\\1\end{array}\right)
$$ regardless of the angle $\alpha$ of the rotation.
This means that a bit flip error (X-error) will change the Grover state from the $|\Psi_+\rangle$ state to the $|\Psi_-\rangle$ state and vice versa. A phase flip (Z error) will also change the Grover state from the $|\Psi_+\rangle$ state to the $|\Psi_-\rangle$ state and vice versa.
In appendix~\ref{sectionOneQubit} we discuss Grover operators on one qubit in more detail.

\subsection{Errors inside Grover Operators}\label{errorsingrovers}
In the preceding sections, we have assumed that errors occur on the Grover register, but outside of the Grover operator. Since most of the circuit complexity is usually within the Grover operator, this assumption may not be a good description of the behavior for real world applications. In the following, we
analyze a simple error model. We focus on
a single-qubit error on the Grover register inside a controlled Grover operator, e.g. an erroneous $X$ gate occurs during the execution
of a controlled Grover operator.

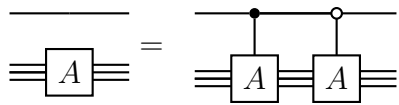
\begin{figure}[H]
\centering
\begin{quantikz}
\qw&\qw&\qw\\
\qwbundle[alternate]{}&\gate{A}\qwbundle[alternate]{}&\qwbundle[alternate]{}
\end{quantikz}
=
\begin{quantikz}
\qw&\ctrl{1}&\octrl{1}\qw&\qw\\
\qwbundle[alternate]{}&\gate{A}\qwbundle[alternate]{}&\gate{A}\qwbundle[alternate]{}&\qwbundle[alternate]{}
\end{quantikz}
\caption{An uncontrolled operation $A$ can be written as the combination of a $0$-controlled and an $1$-controlled operation.}
\label{decompositionA}
\end{figure}

To simplify the following discussion, we start with the observation that an uncontrolled operator
$A$ can be decomposed into a $0$- and an $1$-controlled operator, see fig.~\ref{decompositionA}.

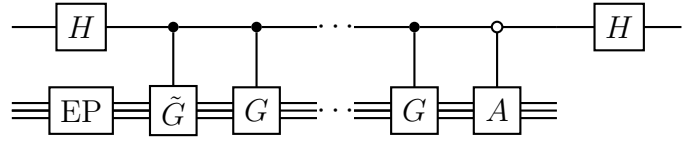
\begin{figure}[H]
\centering
\begin{quantikz}
&\gate{H}              &\ctrl{1}&\ctrl{1}&\qw\cdots&\ctrl{1}&\octrl{1}&\qw&\gate{H}&\qw \\
&\gate{\rm EP}\qwbundle[alternate]{}&\gate{{\tilde G}}\qwbundle[alternate]{}&\gate{G}\qwbundle[alternate]{}&\qwbundle[alternate]{}\cdots&\gate{G}\qwbundle[alternate]{}&\gate{A}\qwbundle[alternate]{}&\qwbundle[alternate]{}
\end{quantikz}
\caption{A controlled Grover operator $G=UV$ with an uncontrolled error $A$ in the first copy of a sequence of kickbacks.
The error $A$ can be split into two controlled operations and the operator with error can be written as controlled
operator ${\tilde G}=UAV$ with an additional $0$-controlled operator $A$ at the end of the sequence of kickbacks.}
\label{errorAtEnd}
\end{figure}

Therefore, if we consider a controlled Grover operator with an uncontrolled single-qubit error in it then we can replace the uncontrolled error
$A$ by the two controlled operators and we can also move the $0$-controlled part to the end of a sequence of kickbacks as it commutes with all $1$-controlled gates, see
fig.~\ref{errorAtEnd}.

If the state in the qubit on the top is in the state
\begin{equation}\label{state}
\frac{1}{\sqrt{2}} \left(\begin{array}{c}1\\ \omega \end{array}\right)=
\frac{1}{\sqrt{2}} \left(\begin{array}{c}1\\ {\rm cos}(\alpha)+i\cdot{\rm sin}(\alpha)\end{array}\right)
\end{equation}
with the phase $\omega={\rm cos}(\alpha)+i\cdot{\rm sin}(\alpha)$ directly before the $H$ gate at the end then we obtain the probability
\begin{equation}\label{stateProbability}
\frac{1-{\rm cos}(\alpha)}{2} = \frac{1-\Re(\omega)}{2}
\end{equation}
for measuring $1$, where $\Re(\omega)$ denotes the real part of a complex number.\footnote{In~\cite{QAE} the
eigenvalues $e^{\pm i 2\theta_a}$ with the corresponding probability ${\rm sin}^2(\theta_a)$ are considered,
this corresponds to $\alpha=2\theta_a$ in our notation and we can use ${\rm cos}(2x)=1-2{\rm sin}^2(x)$ to get the same probability:
$$
\frac{1-{\rm cos}(2\theta_a)}{2}=\frac{1-(1-2{\rm sin}^2(\theta_a))}{2}={\rm sin}^2(\theta_a)
$$}
In the following, we use the partial trace to calculate the probability for measuring $1$. In these cases, the error might lead to an entanglement
between the qubit on the top and the Grover register. Then the calculation of the partial trace is non-trivial and we might interpret the
probability for obtaining the result $1$ as an angle by formula~\ref{stateProbability}.

To calculate the probability of the result $1$ when we have an error $A$ within a controlled Grover operator, we write the
resulting unitary as $G=UV$ for the case without error and the unitary with error as ${\tilde G}=UAV$ as in figure~\ref{errorAtEnd}. Let $|\lambda\rangle$ be an eigenvector
of $G$ that is prepared by the operator ${\rm EP}$, i.e. we have $G|\lambda\rangle=\lambda | \lambda\rangle$. Furthermore, we set
${\tilde G}|\lambda\rangle=|\Phi\rangle$ and $A|\lambda\rangle=|\Psi\rangle$ and we assume that we have $k$ kickbacks
with $G$ on a single qubit and one with ${\tilde G}$.\footnote{It is easy to see that an
error in the last position of a series of kickbacks is the same as one error in the parallel version, which we will introduce in section~\ref{sectionParallelPhaseEstimation}, see for example figure~\ref{LDPhaserCircuit}, as
all the controlled operations in the parallel version commute.}
In the following, we compare an erroneous kickback at the beginning as in fig.~\ref{errorAtEnd} with one at the end.
If the error is in the first kickback then the probability for measuring $1$ is
$$
\frac{1-\Re(\langle \Psi | G^k \Phi\rangle)}{2}
$$
If the error is in the last kickback then the probability for measuring $1$ is
$$
\frac{1-\Re( \lambda^k \langle \Psi | \Phi\rangle)}{2}
$$
In general, if there are $k$ good operators before the operator with error and $\ell$ good operators after it,
then we obtain the result $1$ with the probability
\begin{equation}\label{traceFormulaGeneral}
\frac{1-\Re( \lambda^k \langle \Psi | G^\ell \Phi\rangle)}{2}
\end{equation}

To verify the calculation, we can consider the error-free case if we set $A$ to the identity.
Then there are $k$ good kickbacks and one with the error $I$ leading to
$$
{\rm cos}(\alpha)=\Re( \lambda^k\langle\Psi|\Phi\rangle)=\Re( \lambda^{k+1})
$$
following equations~\ref{state} and~\ref{stateProbability}.
For the phase $\lambda={\rm cos}(b)+i\cdot {\rm sin}(b)$ we can write
$$
\Re(\lambda^{k+1})=\Re(e^{ib(k+1)})=\Re({\rm cos}(b(k+1))+i\cdot {\rm sin}(b(k+1))={\rm cos}(b(k+1))
$$
and this shows that we obtain $k+1$ times the kickback phase $b$ as total angle.\\

It is also instructive to reconsider the case of errors before the Grover operator, i.e. the error model that is used to obtain the majority of the results in this paper, in the light of the partial trace considerations above. An error $A$ before the Grover operator means that we have $\tilde{G}= G A$ and $|\Phi\rangle=GA|\lambda\rangle$ as well as
$|\Psi\rangle=A|\lambda\rangle$. Then
we can decompose $A|\lambda\rangle$ into the following superposition
$$
A|\lambda\rangle=\alpha |\lambda\rangle + \beta | {\overline \lambda}\rangle + \gamma_+ | \mu_+ \rangle
+ \gamma_- | \mu_- \rangle
$$
with orthonormal vectors $|\lambda\rangle$, $|{\overline \lambda}\rangle$, $|\mu_+\rangle$ and $|\mu_-\rangle$ where
$+$ and $-$ correspond to the eigenvalues $+1$ and $-1$. We obtain:
$$
\langle \Psi | \Phi \rangle = |\alpha|^2 \lambda + |\beta|^2{\overline \lambda} + |\gamma_+|^2 - |\gamma_-|^2
$$
For example, if the error $A$ moves $|\lambda\rangle$ completely out of the Grover plane, then we have $\langle \Psi |
\Phi \rangle =  |\gamma_+|^2 - |\gamma_-|^2$. If all eigenvalues are $+1$ then we have $\langle \Psi | \Phi \rangle=1$ and we obtain the correct result up to a missing phase kickback as discussed in section~\ref{secErrorGrover}. As another example, assume that we have
$|\gamma_+| \approx |\gamma_-|$. Then we obtain $\langle \Psi | \Phi \rangle \approx 0$ and this leads to a probability of $0.5$ for the measurement result $1$. In this case, the two eigenvectors are in an equal superposition and the Grover register gets entangled with the qubit that gathers the kickbacks. When we measure only the qubit then the entanglement is destroyed and this leads to the probability of $0.5$. In this case, we cannot obtain
information about $\lambda$.

As final example, assume that $A$ generates an equal superposition of $|\lambda\rangle$ and $|{\overline \lambda}\rangle$, i.e.
we have $|\alpha| \approx |\beta|$ and $\gamma_+,\gamma_-\approx 0$.
If $\lambda$ and ${\overline \lambda}$ are close to $1$ or $-1$ then the magnitude of the scalar product is
also close to $1$, i.e. there is only a small loss of amplitude in the resulting probability of the measurement result $1$. For eigenvectors $\lambda$ and ${\overline \lambda}$ that are close to $i$ and $-i$, respectively, we find that $\langle \Psi | \Phi\rangle$ is close to $0$ if $A$ maps $|\lambda\rangle$ to an equal superposition of both eigenvectors. In this case, for $k$ correct kickbacks and one error on the Grover register before the last kickback we obtain a small value for ${\rm cos}(\alpha)$ in equation~\ref{stateProbability}. This means
that the probability for the result $1$ oscillates around $1/2$ with a small amplitude for an increasing number of kickbacks.

We would like to point out an observation that we made during simulations regarding the eigenvalue spectrum of Grover operators with one bit flip or phase flip errors inside of them. As discussed earlier, one interesting class of Grover operators are those with just one good or just one bad state. For these operators, the eigenvalues of states outside the Grover plane are all +1 or all -1 as discussed in section~\ref{Groveroperators}. We observed that regardless of the operator $M$ we have chosen, the typical outcome was for those eigenvalues to go from all +1 or all -1 to half of them +1 and half of them -1.\\
The implication of this observation is that a single error inside a Grover operator most likely destroys the signal of the QAE. The reason is that the scalar product in the partial trace formula~\ref{traceFormulaGeneral} will most likely contain a superposition of an equal or almost equal number of states with +1 and -1 eigenvalues.

\subsection{Low Depth Amplitude Estimation}\label{StandardLowDepth}
In recent years, the aim to execute amplitude estimation on NISQ (noisy intermediate-scale quantum) devices has led to the development of new techniques. Some of these techniques take the idea of QAE and implement it in such a way that several circuits are executed, where each circuit is not the full circuit in figure~\ref{PE} but only a part of it. 
Furthermore, the QFT of the circuit in figure~\ref{PE} is not executed as a quantum circuit and the results of the individual circuits are combined in a classical post-processing step. The first such algorithm was described in~\cite{AEwithoutPE}, and several alternatives have followed, for example~\cite{BlankLD}, \cite{CQCLD},  \cite{LowDepthAlgos} and~\cite{itQAE}, which are now called low depth QAE\footnote{Another approach to reduce the depth of such circuits is given in \cite{Plekhanov_2022}}. What is relevant for our discussion here is the type of quantum circuit which all of the above methods need to execute. The way the results are processed classically and the way the appropriate circuits are chosen is what distinguishes the different approaches, but this does not concern us here.\\
The structure of quantum circuits to implement such low depth approaches with the aim to get a minimal gate count looks like follows:
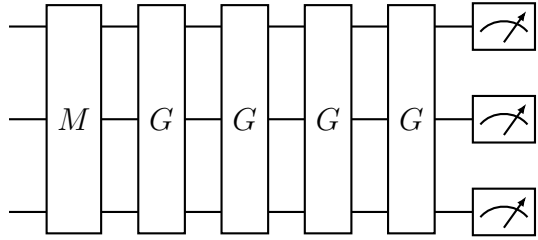
\begin{figure}[H]
\centering
\begin{quantikz}
\qw&\gate[wires=3]{M}  &\gate[wires=3]{G}&\gate[wires=3]{G}&\gate[wires=3]{G}&\gate[wires=3]{G}&\meter{}\\
\qw&  &                 &                 &                 &                 &\meter{}\\
\qw&  &                 &                 &                 &                 &\meter{}
\end{quantikz}
\caption{The circuit for four Grover operators $G$ in a low depth amplitude estimation. The initialization is done with the operator $M$, the model we want to analyze. $M$ creates a superposition of the good states and the bad states, or, equivalently, a superposition of the two eigenstates of the Grover rotation. }
\label{LDQAE}
\end{figure}

In this implementation, we measure the entire register for several runs with a possibly varying number $N$ of Grover operators and we have to count the number of good states among the measured results. The basic idea is that we could plot the number of good states found in the measurements as a function of the number of Grover operators in the circuit and we would see a noisy version of the function ${\rm sin}^2((2 N +1) \theta)$ where $N$ is the number of operators. We can then extract $\theta$ and use the value to calculate the probability $a$ of good states as $a = {\rm sin}^2(\theta)$.
There are various suggestions in the literature that aim to maximize the efficiency of this procedure, by cleverly choosing the number $N$ of operators that we implement in circuits and post-processing the results, see~\cite{AEwithoutPE}, \cite{BlankLD}, \cite{CQCLD}, \cite{LowDepthAlgos} and~\cite{itQAE} and references therein.

Now recall the discussion in section~\ref{secErrorGrover}: An error will destroy the eigenstate in the quantum register, and all subsequent operators will no longer perform a rotation in the Grover plane. So we cannot control the effect of a single error in the Grover register.

This means that this procedure only delivers useful results reliably, when it is very likely that not a single error occurs in the execution of the entire circuit. We can quantify how likely this is: If the probability for an error to occur in an operator is $p$ and the number of operators is $N$ then the probability to measure a meaningful result is $(1-p)^N$. So it drops exponentially in $N$, which means we have to increase the number of measurements exponentially in $N$ to get usable results.\\

We can build an alternative implementation of the idea behind the low depth amplitude estimation if we accept a higher gate count for the circuit by making the operators controlled and collecting the kickbacks. Such circuits can be seen as a part of the circuit in figure~\ref{PE}:

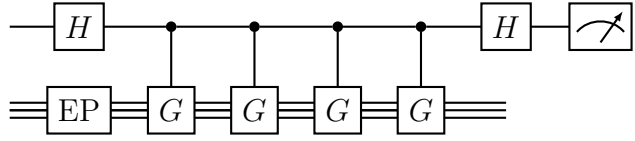
\begin{figure}[H]
\centering
\begin{quantikz}
\qw&\gate{H}              &\ctrl{1}         &\ctrl{1}         &\ctrl{1}         &\ctrl{1}         &\gate{H}&\meter{}\\
\qwbundle[alternate]{}&\gate{\rm EP}\qwbundle[alternate]{}&\gate{G}\qwbundle[alternate]{}&\gate{G}\qwbundle[alternate]{}&\gate{G}\qwbundle[alternate]{}&\gate{G}\qwbundle[alternate]{}&\qwbundle[alternate]{}     & \\
\end{quantikz}
\caption{An alternative implementation of the circuit in figure~\ref{LDQAE}. The low depth amplitude estimation is performed without measuring the Grover register. This circuit generates the state in equation~\ref{equationStateLowDepth} in the top qubit. Here, ${\rm EP}$ is the unitary operator that creates an eigenvector of the Grover operator $G$. The final state is typically measured and the measurements are classically post-processed together with the measurements of circuits implementing other powers of $G$.}
\label{PAEWPE}
\end{figure}

The state of the top qubit in the circuit directly before the measurement is 
\begin{equation}
H \cdot P(2\theta)^N \cdot H \cdot |0\rangle
\label{equationStateLowDepth}
\end{equation}
where $N$ is the number of Grover operators and $P$ is a phase gate. In the example in figure~\ref{PAEWPE} we have $N=4$. Applying the $2\times2$ matrices for the Hadamard gate and the phase gates, it is straightforward to work out the state of the top qubit at the end of the circuit. The probability to obtain the measurement result $1$ for the top qubit when there are $N$ operators in the circuit is the following:

\begin{equation}
\frac{1- {\rm cos}(2N\theta)}{2}= {\rm sin}^2(N \theta) \label{ldOnes}
\end{equation}

To look at a concrete example, we specify a very simple 3-qubit operator $M$ as follows:
\begin{figure}[H]
\centering
\begin{quantikz}
\qw&\gate{U_3(2.86, 0, 0)}&\ctrl{1}               &\qw                    &\qw \\
\qw&\qw                   &\gate{U_3(-2.86, 0, 0)}&\ctrl{1  }             &\qw \\
\qw&\qw                   &\qw                    &\gate{U_3(-2.86, 0, 0)}&\qw
\end{quantikz}
\caption{A simple circuit $M$ which can be used to construct a Grover operator. We use the notation $U_3$ from Qiskit~\cite{qiskit}.}
\label{modelgate}
\end{figure}
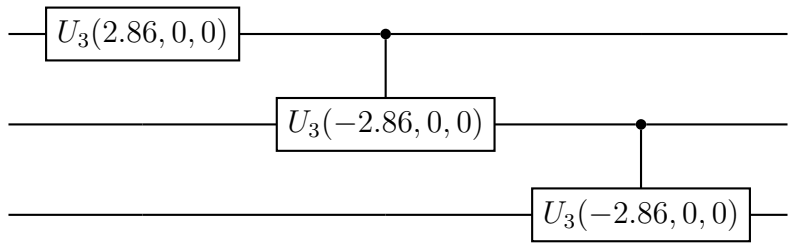
Assume that the good states with non-zero probabilities are $|000\rangle$, $|001\rangle$ and $|011\rangle$. It is now easy to build a Grover operator from $M$ as 
$G= -M S_0M^\dagger S_x$. This Grover operator fulfills the conditions\footnote{The only bad state is $\ket{111}$ and it has a probability of $94,21\%$, all other states are good states. Inspection of the circuit in figure \ref{modelgate} shows, that the remaining good states, e.g. $\ket{100}$, have probability 0 to occur.} set out in section~\ref{secErrorGrover} and therefore has only $+1$ eigenvalues outside the Grover plane, while the other eigenvalues will be close to $+1$. This way, a single error has a limited impact on the resulting kickback.\\
We can now use this Grover operator in circuits like that of figure~\ref{PAEWPE} with an increasing number $N$ of operators $G$ to find the following probabilities for the measurement result 1:

\begin{figure}[H]
\centering
\includegraphics[width=0.8\textwidth]{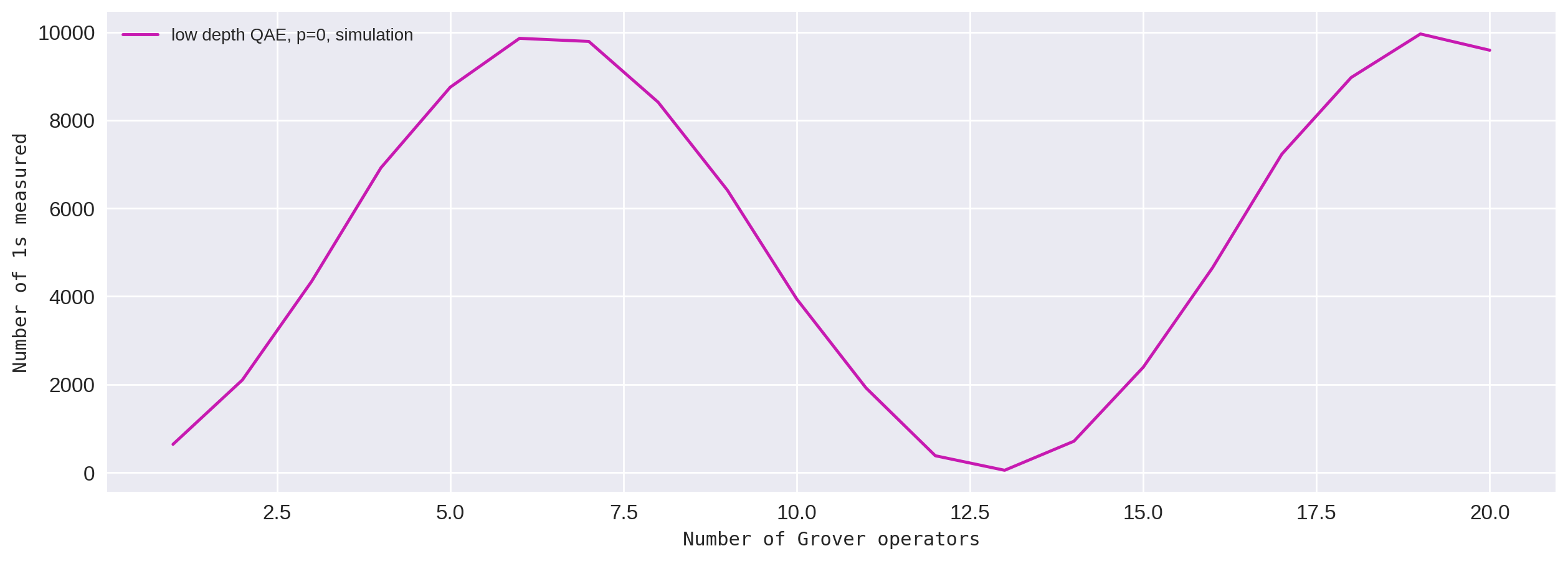}
\caption{The number of ones measured in circuits like in figure~\ref{PAEWPE} as a function of the number of Grover operators. For each number of operators we simulate $10000$ shots.}
\label{plotLDcomparison}
\end{figure}

The information about the probability $a$ of good states is encoded in the wavelength of the signal. The expected number of measurements that is necessary for each circuit depends on the amplitude of the signal. It also depends on the distribution of the measurements around the mean.\\

In our example in figure~\ref{plotLDcomparison}, we see that the first maximum of ${\rm sin}^2(N\theta)$ is close to $N=6$ operator executions, so we have $6\cdot \theta = \pi/2$. Therefore, we calculate $a={\rm sin}^2(\theta) \approx 6\%$ as the probability of the good states.
Of course, the way we read off the information about $\theta$ by plotting the full graph for almost a full wavelength is quite inefficient. The techniques in references~\cite{AEwithoutPE},\cite{BlankLD}, \cite{CQCLD}, \cite{LowDepthAlgos} and~\cite{itQAE} provide ways to determine the probability $a$ with as few executions of the Grover operator as possible.

\subsection{Errors in Amplitude Estimation} \label{errorsinamplitudeestimation}
In the following we discuss the consequences of errors on the Grover register in a sequence of kickbacks of the Grover operator. Errors on the qubits that receive the kickbacks are not modeled. When we construct the circuit for a series of kickbacks we include one error gate in front of each Grover operator with probability $p$. In most examples, this will be an $X$ or a $Z$ gate. These correspond to a bit-flip or a phase flip error on a randomly chosen qubit of the Grover register, respectively. We also considered random unitary operations on all Grover qubits simultaneously in our simulations. The results of those different error models are generally similar.
As discussed in section~\ref{errorsingrovers} we note that our simple error model is not close to reality for most applications.

Suppose an error occurs before one of the Grover operators in standard QAE. In general, the state will be moved out of the Grover plane and the subsequent operator will kick back an eigenvalue of $+1$ or $-1$, or, more generally, a superposition of these. Since the operators are executed serially, not just the first operator after the error is affected, but also all operators that follow kick back the phase $+1$ or $-1$ or a superposition of these.\\

\begin{figure}[H]
\centering
\includegraphics[width=0.8\textwidth]{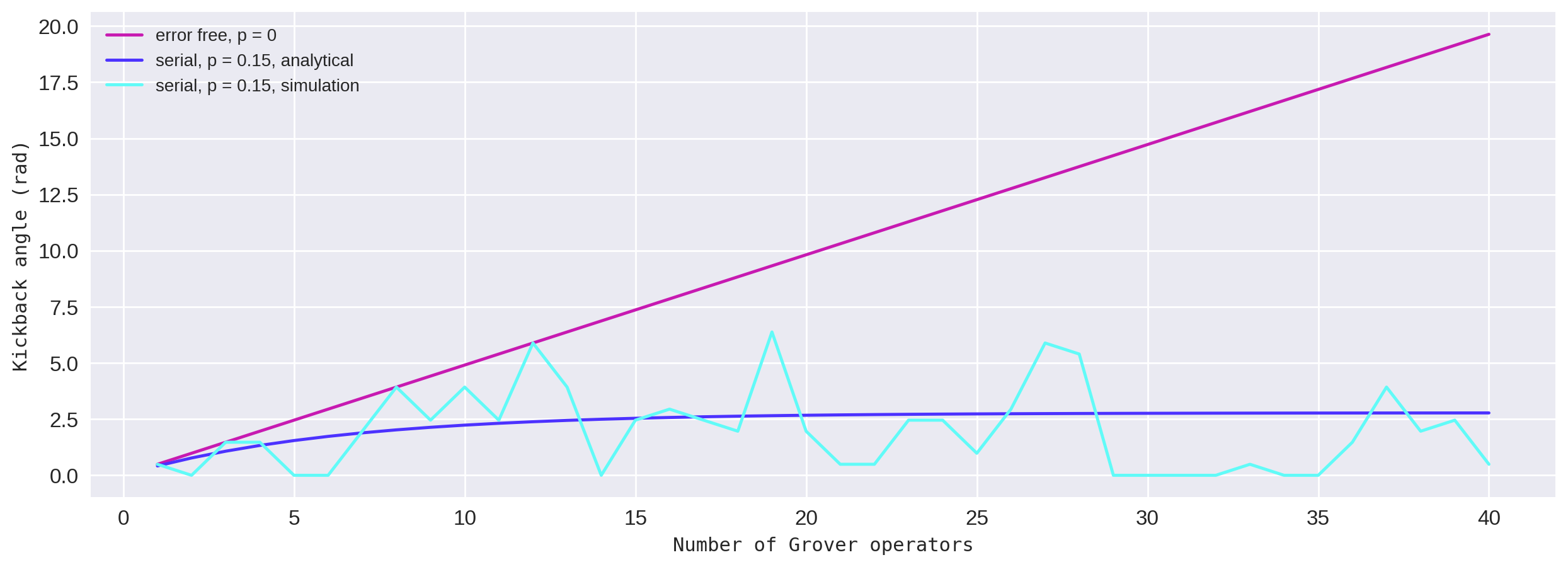}
\caption{The angle kicked back as a function of the number of Grover operators. In the simulation with errors, we plotted a single run for each number of Grover operators, rather than the mean after many runs. The expected value for the serial setup with errors leads to a kickback that is asymptotically constant as a function of the number of Grover operators, which will rule out this approach in practice. The variance of the measurements is large and makes the result even less useful. In contrast, the error free case has a linear dependency.}
\label{plotkickbackangles}
\end{figure}

We consider a sequence of $N$ controlled Grover operators with eigenvalues $e^{\pm 2 i \theta}$ as in figure~\ref{PAEWPE} and the probability $p$ that an error occurs at a given operator. The probability to see an accumulation of $k$ successful kickbacks on the qubit on the top is therefore given by
$$
P(k) = p (1-p)^k \;\;{\rm for}\;\; k<N \quad {\rm and} \quad
P(k) = (1-p)^k \;\;{\rm for}\;\; k=N.
$$
We can describe the number $k$ of cumulative kickbacks as a geometric random variable:
\begin{equation*}
P(k) = p (1-p)^{k}
\end{equation*}
A short calculation shows that for $N>2$, the expected value $2\theta \cdot \mathbb{E}(k)$ for the total kickback  is given by\footnote{This is only true if the error causes $+1$ kickbacks, either because the eigenstate has the eigenvalue $+1$ or because an even number of operators kick back $-1$ each. In cases where an odd number of $-1$ kickbacks occur the results change much more.}
\begin{eqnarray}
\label{kapprox}
2 \theta \cdot \mathbb{E}(k) &=& 2\theta \cdot \left(\frac{1-p}{p} - \frac{(1-p)^{N+1}}{p}\right)
\end{eqnarray}
instead of $2 \theta \cdot \mathbb{E}(k)= 2 N\theta$ in the error free version. For small $N$ and small $p$ this can produce a usable approximation of the linear behavior, but even a 15\% error probability for a Grover operator\footnote{The operator will consist of thousands of gates on dozens of qubits even for modest real world examples.} rules out any practical application.
We can compare the expectation value to the results of a simulation. The results are shown in figure~\ref{plotkickbackangles}.

\begin{figure}[H]
\centering
\includegraphics[width=0.8\textwidth]{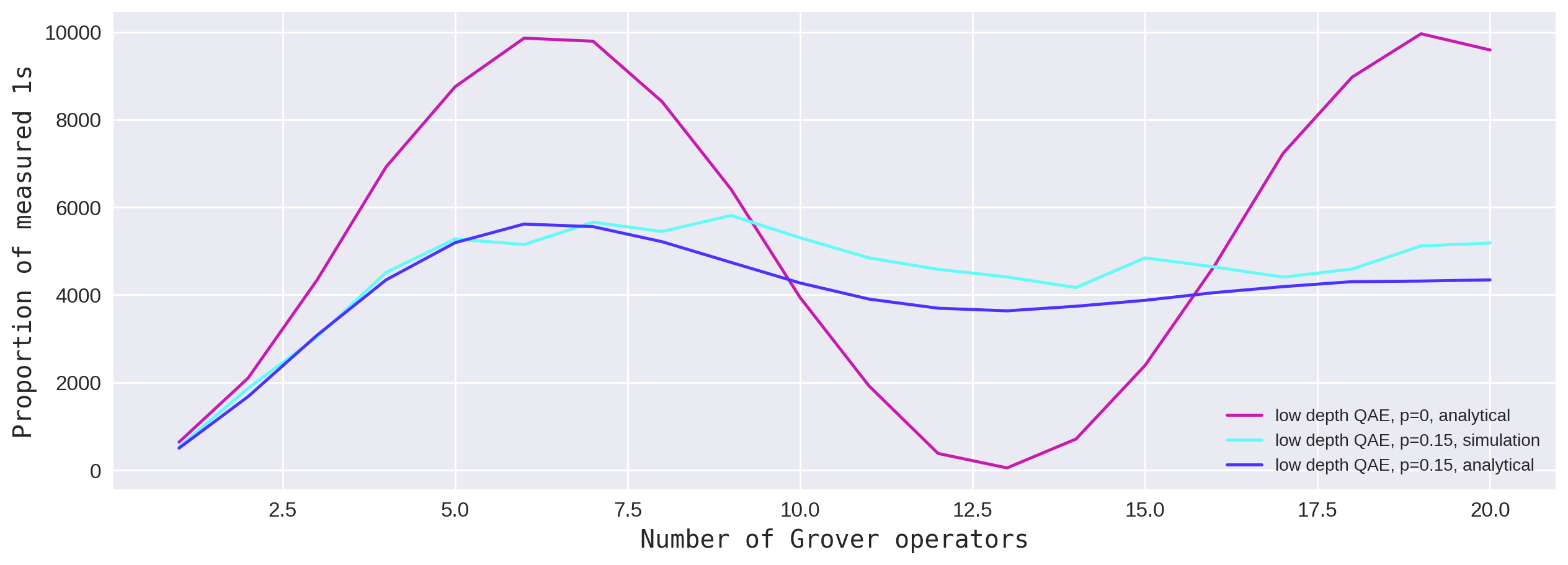}
\caption{The number of times we measure the result 1 in the low depth circuits of figure~\ref{PAEWPE} for an increasing number of operators with 10000 shots per experiment. The probability for a bit flip error to occur on one of the qubits of the Grover register is 15\% for each operator. The cyan line is the result averaged over 100 circuit simulations for each number of Grover operators and the blue line is the theoretical result evaluated numerically.}
\label{plotLDcomparison2}
\end{figure}

Compared to the error free case, this is a major qualitative change in the behavior. Instead of being linear in the number $N$ of Grover operators, the expected value simply converges to a constant value, as is shown in figure~\ref{plotkickbackangles}. This means that an increase of $N$ will have no effect -- not even on average, when we measure the circuit many times. In the error free case of a standard amplitude estimation as in the example of figure~\ref{PE} with three bits precision the three qubits on the top carry the phases  $2\theta$, $4 \theta$ and $8 \theta$ from the kickbacks. In the case of a higher precision, we further have the phases $2^b\theta$ for suitable $b$. In the presence of errors, e.g. with a 15\% error rate, we would have almost the same phase on the 5th and 6th qubit as can be seen in figure~\ref{plotkickbackangles}. The resulting effect is that the information carried by these qubits is lost compared to the error free case. Additionally, the qubits with the smaller kickbacks would also be affected by the errors. So even if a QAE with a precision of 6 qubits is performed, the results will not be better than those of a QAE with 5 qubits. Therefore, the bigger the error rate $p$ is the fewer binary digits in the output of the QAE are meaningful.\\
In the case of low depth QAE as in figure~\ref{PAEWPE}, this means that we do not gain much from executing circuits with more Grover operators than are useful for the error rate. Note that increasing the number of measurements cannot help with this problem because we are looking at expected values\footnote{The number of measurements we need to extract useful information about the expected value depends on the variance of the measurements. Our point here is, that the information carried by the expected value itself stops being useful after a low number of Grover operators. It turns out that the variance makes the situation even worse, since it grows quickly even for a low numbers of operators.}.\\
The probability to measure the result 1 on the top qubit in figure~\ref{PAEWPE} therefore also converges to a constant. If we think about the problem in a continuous setting, which makes sense for large numbers of kickbacks, then the arguments of the trigonometric functions in equation \ref{ldOnes} can be thought of as an exponentially distributed random variable $X$ and the expected value of equation \ref{ldOnes} takes the form
\begin{equation*}
\mathbb{E} \left( \frac{1- {\rm cos}(X)}{2} \right)= \mathbb{E} \left({\rm sin}^2 \left(\frac{X}{2}\right)\right)
\end{equation*}
as long as errors with kickbacks $-1$ do not occur, which is the case by construction for problems that satisfy the requirements set out in section~\ref{secErrorGrover}. To compare the outcomes to the error free case of figure~\ref{plotLDcomparison} we need to calculate the expected value. Unfortunately, this involves a cosine integral and has no expression in terms of elementary functions. Therefore, we calculate the expected value numerically and we find the results that are shown in figure~\ref{plotLDcomparison2}.

The results show that an error probability of 15\% dampens strongly the oscillation of the curve that corresponds to the number of times we measure the result 1 for an increasing number of Grover operators. The curve stays close to its mean value instead of oscillating between $0$ and $1$ as in the error free case. This means that we will need a very large number of measurements to gain useful information about the angle $\theta$. This is already caused by the mean value. The variance of the measurement results makes the situation even worse.
From equation~\ref{kapprox}, we can read off a ceiling for the number $M$ of kickbacks that is expected to be generated when we execute a large number $N$ of Grover operators: $M = (1-p)/p$.\\
The reason for the poor performance of the circuits with noise is the fact that the oracles are executed serially and our errors do not commute with the Grover operators. A serial low depth QAE with noise was also analyzed in~\cite{QAEwithoutPEnoise}. The results looked more promising than ours since the error is modeled as depolarizing noise, which commutes with all gates in the circuit. Note that the consequences of commuting errors in serial circuits are essentially the same as in parallel circuits, which will be defined and analyzed in the following sections.

\section{Parallelization of Phase Estimation}\label{sectionParallelPhaseEstimation}
In the previous section, we analyzed different approaches to QAE and we showed that they are very sensitive to errors -- to such an extent, that real life applications seem to be out of reach for the foreseeable future \footnote{Suppose that a problem, which would be hard on a classical computer, has to be analyzed with an accuracy of 0.1\%. This corresponds to a QAE with a 10-qubit resolution and so approximately 1000 Grover operators are executed in a circuit as in figure~\ref{PE}. Let us assume that we can implement the model on 100 qubits with 10 gates acting on each qubit. The corresponding Grover operator contains the model circuit and its inverse, contributing 2000 gates. Marking the 0-state and the good states will require gates with approximately 100 controls, each of which we might manage to decompose into roughly 2000 gates with single controls and even more uncontrolled gates. So our optimistic estimate is that at least 10000 gates are required to implement such a Grover operator. Therefore, our circuit will contain at least 10 million gates,  which we need to execute with a high probability of not incurring a single error. This leads to the rough estimate that a 99.99999\% fidelity will be necessary to deal with such a problem.}. Therefore, it is necessary to find techniques that improve QAE on noisy hardware and we propose the 
parallelization of the underlying phase estimation as one promising approach.

In the following, we describe three ways to parallelize a phase estimation: A parallelization with additional registers to create the kickbacks, a parallelization with entangled qubits to collect kickbacks, and a parallelization with intermediate resets of qubits. Since the amplitude estimation is a special case of phase estimation with the Grover operator as unitary operator, we can also apply these techniques to amplitude estimation. This is discussed in more detail in section~\ref{sectionParalleQAE}.

\subsection{Simple Parallelization}\label{subsectionSimpleParallel}
Assume that the standard quantum phase estimation algorithm estimates the eigenvalues of an operator $G$ in binary encoding with $b$ bits precision. The algorithm requires $2^b-1$ copies of $G$ which are executed consecutively on the same quantum register as shown in figure~\ref{PE}.

The parallelization of this circuit is achieved by introducing $2^b-2$ new quantum registers, one for each operator $G$. The circuit of the example with $b=3$ bit then looks like follows:
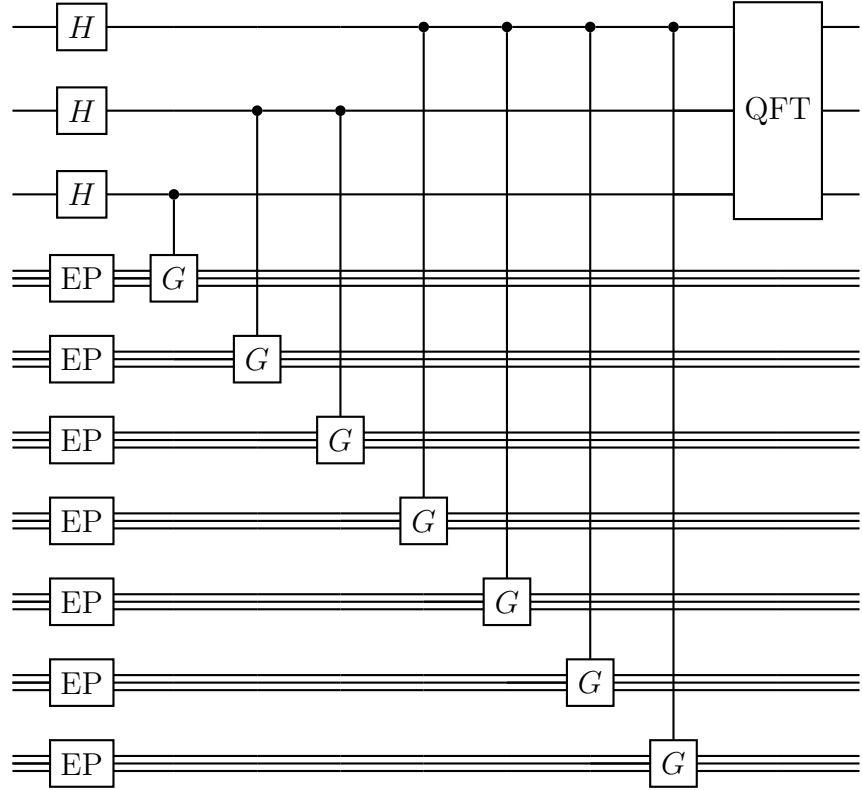
\begin{figure}[H]
\centering
\begin{quantikz}
\qw & \gate{H}      & \qw      & \qw      & \qw      & \ctrl{6} & \ctrl{7} & \ctrl{8} & \ctrl{9}& \gate[wires=3]{\rm QFT}& \qw \\
\qw & \gate{H}      & \qw      & \ctrl{3} & \ctrl{4} & \qw      & \qw      & \qw      & \qw     & \qw                    & \qw \\
\qw & \gate{H}      & \ctrl{1} & \qw      & \qw      & \qw      & \qw      & \qw      & \qw     & \qw                    & \qw \\
\qwbundle[alternate]{} & \gate{\rm EP}\qwbundle[alternate]{} & \gate{G}\qwbundle[alternate]{} & \qwbundle[alternate]{}      & \qwbundle[alternate]{}      & \qwbundle[alternate]{}      & \qwbundle[alternate]{}      & \qwbundle[alternate]{}      & \qwbundle[alternate]{}     & \qwbundle[alternate]{}                    & \qwbundle[alternate]{} \\
\qwbundle[alternate]{} & \gate{\rm EP}\qwbundle[alternate]{} & \qwbundle[alternate]{}      & \gate{G}\qwbundle[alternate]{} & \qwbundle[alternate]{}      & \qwbundle[alternate]{}      & \qwbundle[alternate]{}      & \qwbundle[alternate]{}      & \qwbundle[alternate]{}     & \qwbundle[alternate]{}                    & \qwbundle[alternate]{} \\
\qwbundle[alternate]{} & \gate{\rm EP}\qwbundle[alternate]{} & \qwbundle[alternate]{}      & \qwbundle[alternate]{}      & \gate{G}\qwbundle[alternate]{} & \qwbundle[alternate]{}      & \qwbundle[alternate]{}      & \qwbundle[alternate]{}      & \qwbundle[alternate]{}     & \qwbundle[alternate]{}                    & \qwbundle[alternate]{} \\
\qwbundle[alternate]{} & \gate{\rm EP}\qwbundle[alternate]{} & \qwbundle[alternate]{}      & \qwbundle[alternate]{}      & \qwbundle[alternate]{}      & \gate{G}\qwbundle[alternate]{} & \qwbundle[alternate]{}      & \qwbundle[alternate]{}      & \qwbundle[alternate]{}     & \qwbundle[alternate]{}                    & \qwbundle[alternate]{} \\
\qwbundle[alternate]{} & \gate{\rm EP}\qwbundle[alternate]{} & \qwbundle[alternate]{}      & \qwbundle[alternate]{}      & \qwbundle[alternate]{}      & \qwbundle[alternate]{}      & \gate{G}\qwbundle[alternate]{} & \qwbundle[alternate]{}      & \qwbundle[alternate]{}     & \qwbundle[alternate]{}                    & \qwbundle[alternate]{} \\
\qwbundle[alternate]{} & \gate{\rm EP}\qwbundle[alternate]{} & \qwbundle[alternate]{}      & \qwbundle[alternate]{}      & \qwbundle[alternate]{}      & \qwbundle[alternate]{}      & \qwbundle[alternate]{}      & \gate{G}\qwbundle[alternate]{} & \qwbundle[alternate]{}     & \qwbundle[alternate]{}                    & \qwbundle[alternate]{} \\
\qwbundle[alternate]{} & \gate{\rm EP}\qwbundle[alternate]{} & \qwbundle[alternate]{}      & \qwbundle[alternate]{}      & \qwbundle[alternate]{}      & \qwbundle[alternate]{}      & \qwbundle[alternate]{}      & \qwbundle[alternate]{}      & \gate{G}\qwbundle[alternate]{}& \qwbundle[alternate]{}                    & \qwbundle[alternate]{}
\end{quantikz}
\caption{A parallel version of the quantum phase estimation circuit in figure~\ref{PE}. The unitary EP generates a single eigenstate of $G$.}
\label{PPE}
\end{figure}
For each copy of the register we need an additional EP gate and it is necessary that EP generates a single eigenstate of the operator $G$ only and not a superposition of eigenstates. The kickbacks can be done in parallel for the qubits that go into the QFT. As the qubit on the top has $2^{b-1}$ kickbacks this qubit is the bottleneck of this parallelization.

\subsection{Indirect Parallelization}\label{secIndParallel}
An alternative way to achieve a parallelization is based on entanglement. It requires one extra qubit for each operator $G$ and it achieves a higher level of parallelization than the method of section~\ref{subsectionSimpleParallel}. The main idea is to replace the sequence of kickbacks on a qubit by parallel kickbacks on entangled qubits. The execution of all controlled $G$ operators is truly parallel, that is, the gate depth of the circuit shrinks to essentially the execution of a single controlled $G$ operator.\\

For example, two kickbacks with the operator $G$ on a qubit in the state $H|0\rangle$ can be replaced by two parallel executions of $G$ as follows:
\begin{figure}[H]
\centering
\begin{quantikz}
\qw & \gate{H}     & \ctrl{2} & \ctrl{1} & \ctrl{2}& \qw \\
\qwbundle[alternate]{} & \gate{\rm EP}\qwbundle[alternate]{}& \qwbundle[alternate]{}      & \gate{G}\qwbundle[alternate]{} & \qwbundle[alternate]{}     & \qwbundle[alternate]{} \\
\qw & \qw          & \targ{}  & \ctrl{1} & \targ{} & \qw \\
\qwbundle[alternate]{} & \gate{\rm EP}\qwbundle[alternate]{}& \qwbundle[alternate]{}      & \gate{G}\qwbundle[alternate]{} & \qwbundle[alternate]{}     & \qwbundle[alternate]{}
\end{quantikz}
\caption{A circuit that leads to the same kickback on the first qubit as two serial kickbacks. The two copies of $G$ can now be executed fully simultaneously. The unitary ${\rm EP}$ generates a single eigenvector of $G$.}
\label{EPEsimple}
\end{figure}
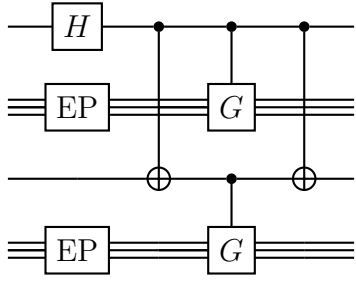
Here, instead of kicking back the phases directly to one qubit, each operator $G$ kicks back the phase to a different qubit, such that each operation $G$ is fully independent of the execution of any other $G$. The total phase kickback is obtained by entangling the qubits before the kickbacks and by untangling the qubits after the kickbacks.

To see how this principle works, we look at the state of the first and third qubits from the top in figure \ref{EPEsimple}. After the execution of the $H$ gate and the first CNOT  gate, these two qubits are in the entangled state:
$$
\frac{1}{\sqrt{2}}\left( |00\rangle + |11\rangle\right)
$$
The idea now is that the entangled state obtains one phase for $|11\rangle$ from each of the two $G$ operators, adding up to the square of the phase, i.e.,
the resulting state is
$$
\frac{1}{\sqrt{2}}\left( |00\rangle + \lambda^2 |11\rangle\right)
$$
where $\lambda$ denotes the unknown eigenvalue of $G$, i.e., $G|\lambda\rangle=\lambda|\lambda\rangle$. The final CNOT gate leads to the state
$$
\frac{1}{\sqrt{2}}\left( |0\rangle + \lambda^2 |1\rangle\right)\otimes |0\rangle
$$
and the state of the first qubit corresponds to the state that is obtained by the standard phase estimation algorithm for the operator $G^2$.

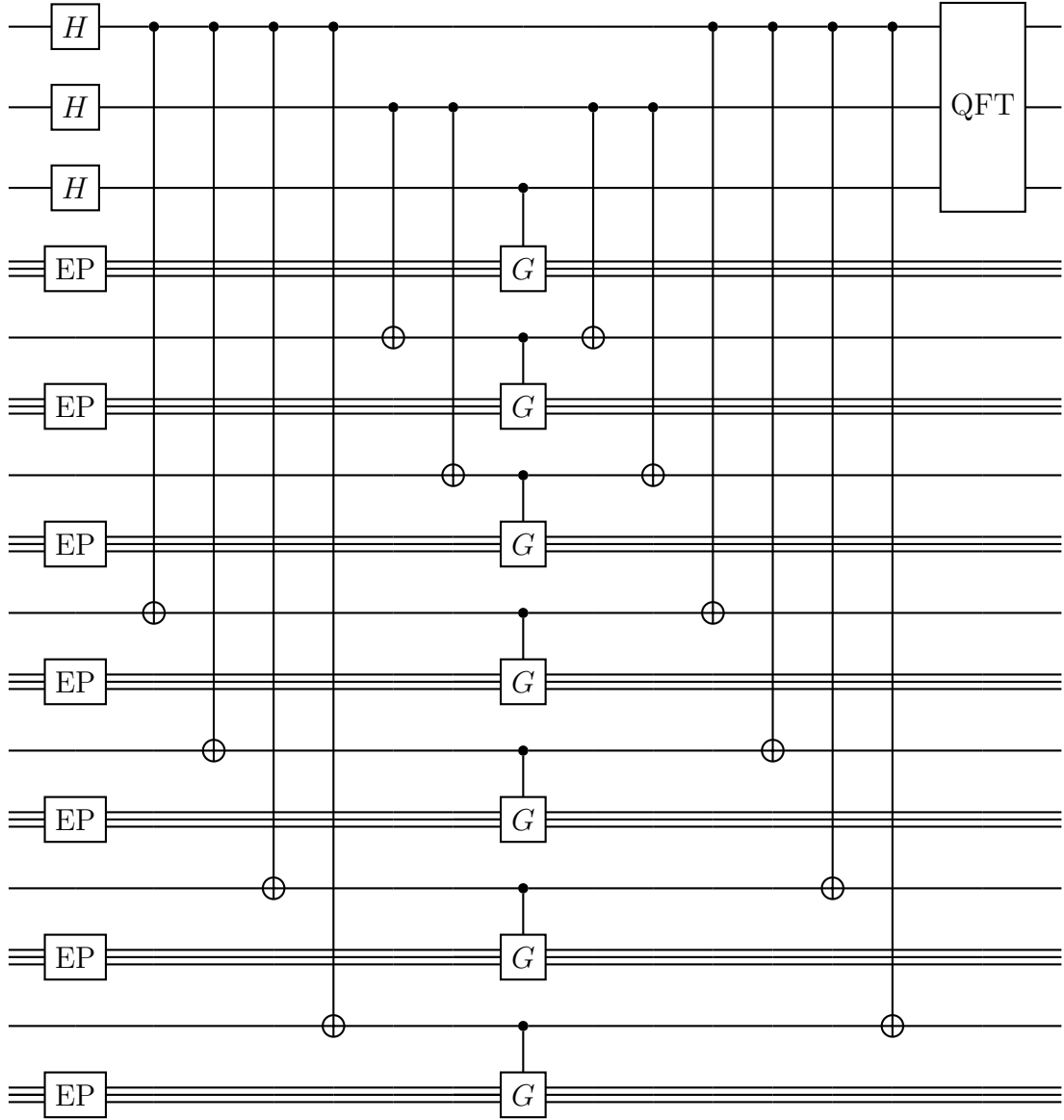
\begin{figure}[H]
\centering
\begin{quantikz}
\qw & \gate{H}       & \ctrl{8} & \ctrl{10} & \ctrl{12} & \ctrl{14} & \qw     & \qw     & \qw      & \qw     & \qw     & \ctrl{8} & \ctrl{10} & \ctrl{12} & \ctrl{14} &  \gate[wires=3]{\rm QFT}    & \qw \\
\qw & \gate{H}       & \qw      & \qw       & \qw       & \qw       & \ctrl{3}& \ctrl{5}& \qw      & \ctrl{3}& \ctrl{5}& \qw      & \qw       & \qw       & \qw       &                             & \qw \\
\qw & \gate{H}       & \qw      & \qw       & \qw       & \qw       & \qw     & \qw     & \ctrl{1} & \qw     & \qw     & \qw      & \qw       & \qw       & \qw       &                             & \qw \\
\qwbundle[alternate]{} & \gate{\rm EP}\qwbundle[alternate]{}  & \qwbundle[alternate]{}      & \qwbundle[alternate]{}       & \qwbundle[alternate]{}       & \qwbundle[alternate]{}       & \qwbundle[alternate]{}     & \qwbundle[alternate]{}     & \gate{G}\qwbundle[alternate]{} & \qwbundle[alternate]{}     & \qwbundle[alternate]{}     & \qwbundle[alternate]{}      & \qwbundle[alternate]{}       & \qwbundle[alternate]{}       & \qwbundle[alternate]{}       &  \qwbundle[alternate]{}                        & \qwbundle[alternate]{} \\
\qw & \qw            & \qw      & \qw       & \qw       & \qw       & \targ{} & \qw     & \ctrl{1} & \targ{} & \qw     & \qw      & \qw       & \qw       & \qw       &  \qw                        & \qw \\
\qwbundle[alternate]{} & \gate{\rm EP}\qwbundle[alternate]{}  & \qwbundle[alternate]{}      & \qwbundle[alternate]{}       & \qwbundle[alternate]{}       & \qwbundle[alternate]{}       & \qwbundle[alternate]{}     & \qwbundle[alternate]{}     & \gate{G}\qwbundle[alternate]{} & \qwbundle[alternate]{}     & \qwbundle[alternate]{}     & \qwbundle[alternate]{}      & \qwbundle[alternate]{}       & \qwbundle[alternate]{}       & \qwbundle[alternate]{}       &  \qwbundle[alternate]{}                        & \qwbundle[alternate]{} \\
\qw & \qw            & \qw      & \qw       & \qw       & \qw       & \qw     & \targ{} & \ctrl{1} & \qw     & \targ{} & \qw      & \qw       & \qw       & \qw       &  \qw                        & \qw \\
\qwbundle[alternate]{} & \gate{\rm EP}\qwbundle[alternate]{}  & \qwbundle[alternate]{}      & \qwbundle[alternate]{}       & \qwbundle[alternate]{}       & \qwbundle[alternate]{}       & \qwbundle[alternate]{}     & \qwbundle[alternate]{}     & \gate{G}\qwbundle[alternate]{} & \qwbundle[alternate]{}     & \qwbundle[alternate]{}     & \qwbundle[alternate]{}      & \qwbundle[alternate]{}       & \qwbundle[alternate]{}       & \qwbundle[alternate]{}       &  \qwbundle[alternate]{}                        & \qwbundle[alternate]{} \\
\qw & \qw            & \targ{}  & \qw       & \qw       & \qw       & \qw     & \qw     & \ctrl{1} & \qw     & \qw     & \targ{}  & \qw       & \qw       & \qw       &  \qw                        & \qw \\
\qwbundle[alternate]{} & \gate{\rm EP}\qwbundle[alternate]{}  & \qwbundle[alternate]{}      & \qwbundle[alternate]{}       & \qwbundle[alternate]{}       & \qwbundle[alternate]{}       & \qwbundle[alternate]{}     & \qwbundle[alternate]{}     & \gate{G}\qwbundle[alternate]{} & \qwbundle[alternate]{}     & \qwbundle[alternate]{}     & \qwbundle[alternate]{}      & \qwbundle[alternate]{}       & \qwbundle[alternate]{}       & \qwbundle[alternate]{}       &  \qwbundle[alternate]{}                        & \qwbundle[alternate]{} \\
\qw & \qw            & \qw      & \targ{}   & \qw       & \qw       & \qw     & \qw     & \ctrl{1} & \qw     & \qw     & \qw      & \targ{}   & \qw       & \qw       &  \qw                        & \qw \\
\qwbundle[alternate]{} & \gate{\rm EP}\qwbundle[alternate]{}  & \qwbundle[alternate]{}      & \qwbundle[alternate]{}       & \qwbundle[alternate]{}       & \qwbundle[alternate]{}       & \qwbundle[alternate]{}     & \qwbundle[alternate]{}     & \gate{G}\qwbundle[alternate]{} & \qwbundle[alternate]{}     & \qwbundle[alternate]{}     & \qwbundle[alternate]{}      & \qwbundle[alternate]{}       & \qwbundle[alternate]{}       & \qwbundle[alternate]{}       &  \qwbundle[alternate]{}                        & \qwbundle[alternate]{} \\
\qw & \qw            & \qw      & \qw       & \targ{}   & \qw       & \qw     & \qw     & \ctrl{1} & \qw     & \qw     & \qw      & \qw       & \targ{}   & \qw       &  \qw                        & \qw \\
\qwbundle[alternate]{} & \gate{\rm EP}\qwbundle[alternate]{}  & \qwbundle[alternate]{}      & \qwbundle[alternate]{}       & \qwbundle[alternate]{}       & \qwbundle[alternate]{}       & \qwbundle[alternate]{}     & \qwbundle[alternate]{}     & \gate{G}\qwbundle[alternate]{} & \qwbundle[alternate]{}     & \qwbundle[alternate]{}     & \qwbundle[alternate]{}      & \qwbundle[alternate]{}       & \qwbundle[alternate]{}       & \qwbundle[alternate]{}       &  \qwbundle[alternate]{}                        & \qwbundle[alternate]{} \\
\qw & \qw            & \qw      & \qw       & \qw       & \targ{}   & \qw     & \qw     & \ctrl{1} & \qw     & \qw     & \qw      & \qw       & \qw       & \targ{}   &  \qw                        & \qw \\
\qwbundle[alternate]{} & \gate{\rm EP}\qwbundle[alternate]{}  & \qwbundle[alternate]{}      & \qwbundle[alternate]{}       & \qwbundle[alternate]{}       & \qwbundle[alternate]{}       & \qwbundle[alternate]{}     & \qwbundle[alternate]{}     & \gate{G}\qwbundle[alternate]{} & \qwbundle[alternate]{}     & \qwbundle[alternate]{}     & \qwbundle[alternate]{}      & \qwbundle[alternate]{}       & \qwbundle[alternate]{}       & \qwbundle[alternate]{}       &  \qwbundle[alternate]{}                        & \qwbundle[alternate]{} \\
\end{quantikz}
\caption{A circuit for a phase estimation with three bits precision, that is parallelized with the entanglement based method.}
\label{EPE}
\end{figure}

The generalization to higher powers of $G$ is straightforward and we can use a
parallelized version of the kickbacks for
$$
G, G^2, G^4, \ldots, G^{2^{b-1}}
$$
in a phase estimation circuit with $b$ bits of precision.  An example for such a circuit
is shown in figure~\ref{EPE}, which is an alternative to the circuit in figure~\ref{PPE}.

It can easily be seen that the maximum gate depth before the QFT operation arises in the top qubit of this construction. The maximum depth $D_{\rm parallel}$ is
\begin{equation}
D_{\rm parallel} = 1+2^b+d(G)
\end{equation}
where $b$ is the number of output qubits and $d(G)$ is the gate depth of the circuit implementation of the controlled $G$ gate. The standard QPE algorithm, for example see ref.~\cite{NC}, has a gate depth $D_{\rm serial}$ of
\begin{equation}
D_{\rm serial} = 1+2^{b-1} d(G).
\end{equation}
The factor by which our method lowers the gate depth is therefore given by
\begin{equation}
\frac{D_{\rm parallel}}{D_{\rm serial}} \approx \frac{1}{2^{b-1}}
\end{equation}
For example, this shows that even for a relatively low resolution of $b=8$ the gate depth of the circuit is reduced by a factor of 128. Of course, this comes at the cost of requiring more qubits, by approximately the same factor.\\
As the parallelization works for all powers of $G$ the method is also useful for low depth versions of QAE, see section~\ref{sectionParallelLowDepthQAE}.
In this case, the relevant comparison is between the most complex circuit we have to construct, i.e. the circuit with the highest number of Grover operators. Depending on the version of low depth QAE, the improvement in depth from parallelization is then either the same as for standard QAE (for example in the case of~\cite{AEwithoutPE}), or less for those approaches where a reduction of circuit depth is traded for more executions of Grover operators (for example in the case of~\cite{LowDepthAlgos}). 

To keep the constructions and drawings of circuits in the following parts of this paper as simple and instructive as possible, we use the parallelization of section~\ref{subsectionSimpleParallel} when we consider parallel versions of phase estimation. It is possible to replace this method with the other methods that we present in this paper.
Furthermore, it is also possible to construct circuits that make use of the different methods at the same time, e.g., we could split a serial sequence of kickbacks into two sequences of parallelized kickbacks as in section~\ref{subsectionSimpleParallel} and executed both sequences in parallel with the entanglement based method.

\subsection{Parallelization by Reinitialization}\label{secReinit}
\begin{figure}[H]
\centering
\begin{quantikz}
& \gate{H} & \qw                       & \qw                    & \qw               & \qw                       & \qw                    & \ctrl{2}          & \qw                       & \qw                    & \ctrl{2}        &\gate[wires=2]{\rm QFT}  &\qw \\
& \gate{H} & \qw                       & \qw                    & \ctrl{1}          & \qw                       & \qw                    & \qw               & \qw                       & \qw                    & \qw               &&\qw \\
& \qwbundle[alternate]{}           & \gate{|0\rangle}\qwbundle[alternate]{} & \gate{\rm EP}\qwbundle[alternate]{} & \gate{G}\qwbundle[alternate]{} & \gate{|0\rangle}\qwbundle[alternate]{} & \gate{\rm EP}\qwbundle[alternate]{} & \gate{G}\qwbundle[alternate]{} & \gate{|0\rangle}\qwbundle[alternate]{} & \gate{\rm EP}\qwbundle[alternate]{} & \gate{G}\qwbundle[alternate]{} &\qwbundle[alternate]{}
\end{quantikz}
\caption{A QPE with two qubits precision with the alternating scheme of reinitialization and kickbacks for a unitary operator $G$, where ${\rm EP}$ prepares an eigenstate $|\lambda\rangle$ of $G$. The gate $|0\rangle$ is a non-unitary operation that reinitializes the qubit to the state $|0\rangle$.}
\label{PE2}
\end{figure}
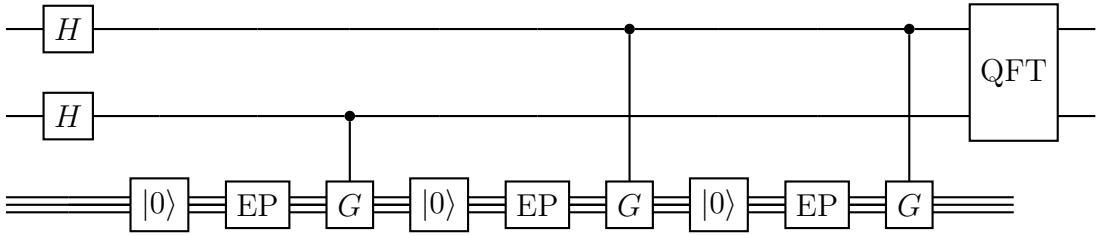

Currently, the limited availability of physical qubits restricts the implementation of parallel QPE and QAE on real hardware. The situation for simulations is similar as it is not possible to simulate much more than 30 qubits on classical computers. A way out is to reset the register of an operator $G$ after creating a kickback. Then the register is in the state $|0\rangle$ and it can be used for another kickback after creating the eigenstate again. The initialization of a qubit is a non-unitary operator and can be done on several hardware types during a calculation~\cite{ibmreset}. An example with three kickbacks is shown in figure~\ref{PE2}.

Of course, with this scheme, we give up the advantages of a low gate depth, i.e., this kind of parallelization does not lead to shorter run times on hardware and we cannot expect to reduce decoherence. On the contrary, we even increase the run time since we have additional initializations and eigenvector preparations compared to standard QPE. However, the reduced number of qubits allows us to simulate the behavior of parallelized versions of QPE and QAE for bigger operators and for more bits of precision because the method with resets and the methods with additional qubits lead to the same results in the error free case while requiring much less memory for the simulation.

\section{Parallel Amplitude Estimation}\label{sectionParalleQAE}
As pointed out in section~\ref{sectionStandardPhaseEstimation}, amplitude estimation is a special case of phase estimation, where the operator $G$ is a Grover operator. Therefore, the parallelization of phase estimation as described in section~\ref{sectionParallelPhaseEstimation} provides the circuits for parallel amplitude estimation as well.  However, note that in the standard amplitude estimation circuit the initialization EP may create a suitable superposition of eigenvalues and this superposition still leads to a unique result. In the case of parallel circuits, the gate EP has to create a single eigenstate and not a superposition, because otherwise the kickbacks for different eigenvalues interfere and this destroys the result.

In this section we discuss some properties of the parallel QAE. We start with the low depth version and we analyze the effect of errors on the result.

\subsection{Parallel Low Depth Amplitude Estimation}\label{sectionParallelLowDepthQAE}
The parallelization of the low depth amplitude estimation as described in section ~\ref{StandardLowDepth} can be achieved with any of the methods in section~\ref{sectionParallelPhaseEstimation}. To illustrate how such circuits look, we choose the method from section~\ref{subsectionSimpleParallel} and we obtain a circuit that looks as follows:

\begin{figure}[H]
\centering
\begin{quantikz}
&\gate{H}      & \ctrl{1} & \ctrl{2}& \ctrl{3}&\ctrl{4} & \gate{H} & \qw \\
&\gate{\rm EP}\qwbundle[alternate]{} & \gate{G}\qwbundle[alternate]{} & \qwbundle[alternate]{}     & \qwbundle[alternate]{}     &\qwbundle[alternate]{}      & \qwbundle[alternate]{}      & \qwbundle[alternate]{} \\
&\gate{\rm EP}\qwbundle[alternate]{} & \qwbundle[alternate]{}      & \gate{G}\qwbundle[alternate]{}& \qwbundle[alternate]{}     &\qwbundle[alternate]{}      & \qwbundle[alternate]{}      & \qwbundle[alternate]{} \\
&\gate{\rm EP}\qwbundle[alternate]{} & \qwbundle[alternate]{}      & \qwbundle[alternate]{}     & \gate{G}\qwbundle[alternate]{}&\qwbundle[alternate]{}      & \qwbundle[alternate]{}      & \qwbundle[alternate]{} \\
&\gate{\rm EP}\qwbundle[alternate]{} & \qwbundle[alternate]{}      & \qwbundle[alternate]{}     & \qwbundle[alternate]{}     &\gate{G}\qwbundle[alternate]{} & \qwbundle[alternate]{}      & \qwbundle[alternate]{}
\end{quantikz}
\caption{Circuit for a parallel low depth amplitude estimation. The gate $G$ is a Grover operator and each quantum register has to be initialized with the same eigenvector $|\lambda\rangle$ of $G$ by the EP gate.}
\label{LDPhaserCircuit}
\end{figure}
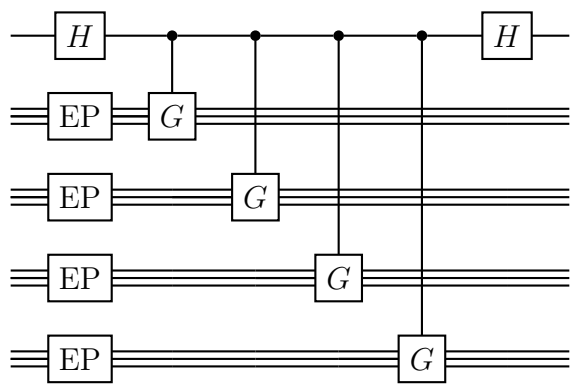

\subsection{Errors in Parallel Low Depth Amplitude Estimation} \label{errorsinldqae}
In this section, we analyze the behavior of the parallel version of amplitude estimation when there
are errors before Grover operators but not inside the operators. We demonstrate an advantages over the corresponding serial version. We focus on low depth circuits, because all operators $G$ kick the phase back onto the same qubit and this makes the analysis simpler. Note that the angle of one kickback is $2\theta$ as mentioned in section~\ref{Groveroperators}. The corresponding calculation for the serial amplitude estimation is in section~\ref{errorsinamplitudeestimation}.\\

\begin{figure}[H]
\centering
\includegraphics[width=0.8\textwidth]{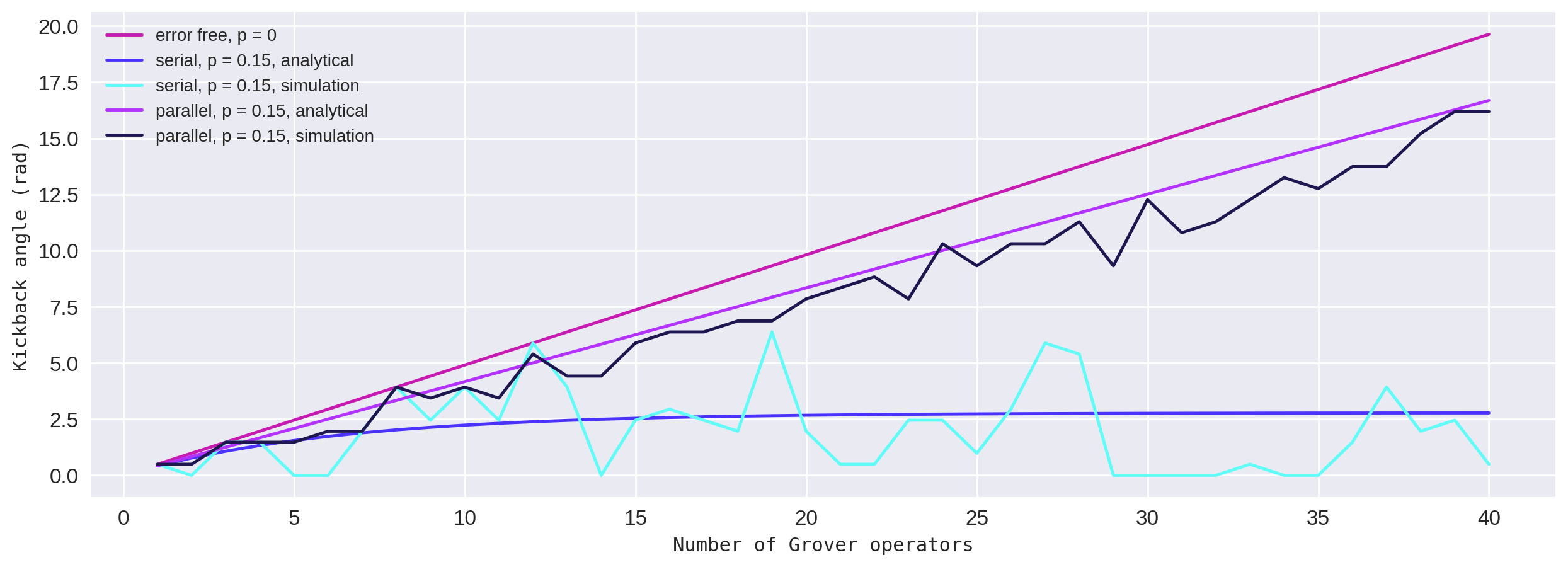}
\caption{The expected angle kicked back as a function of the number of Grover operators. The parallel setup leads to a kickback which is a linear function of the number of Grover operators. The results for the circuit simulations show the values for a single instance for each number of oracles instead of the mean over many runs.}
\label{plotkickbackangles2}
\end{figure}

Let $k$ be the number of error free kickbacks of $N$ parallel operators onto one qubit in figure~\ref{LDPhaserCircuit} before the second $H$ gate. Then $k$ is described by a binomial distribution. This immediately gives us the expectation value for the total angle $2\theta k$ of the kickbacks when there are $N$ operators and the error probability for an operator is $p$:
\begin{equation}
\mathbb{E}(2\theta k) = 2 \theta (1-p) N \label{parallelkicks}
\end{equation}
In contrast, the expectation value for the case without errors is $\mathbb{E}( 2 \theta k) = 2 \theta N$. This is illustrated in figure \ref{plotkickbackangles2}.

\begin{figure}[H]
\centering
\includegraphics[width=0.8\textwidth]{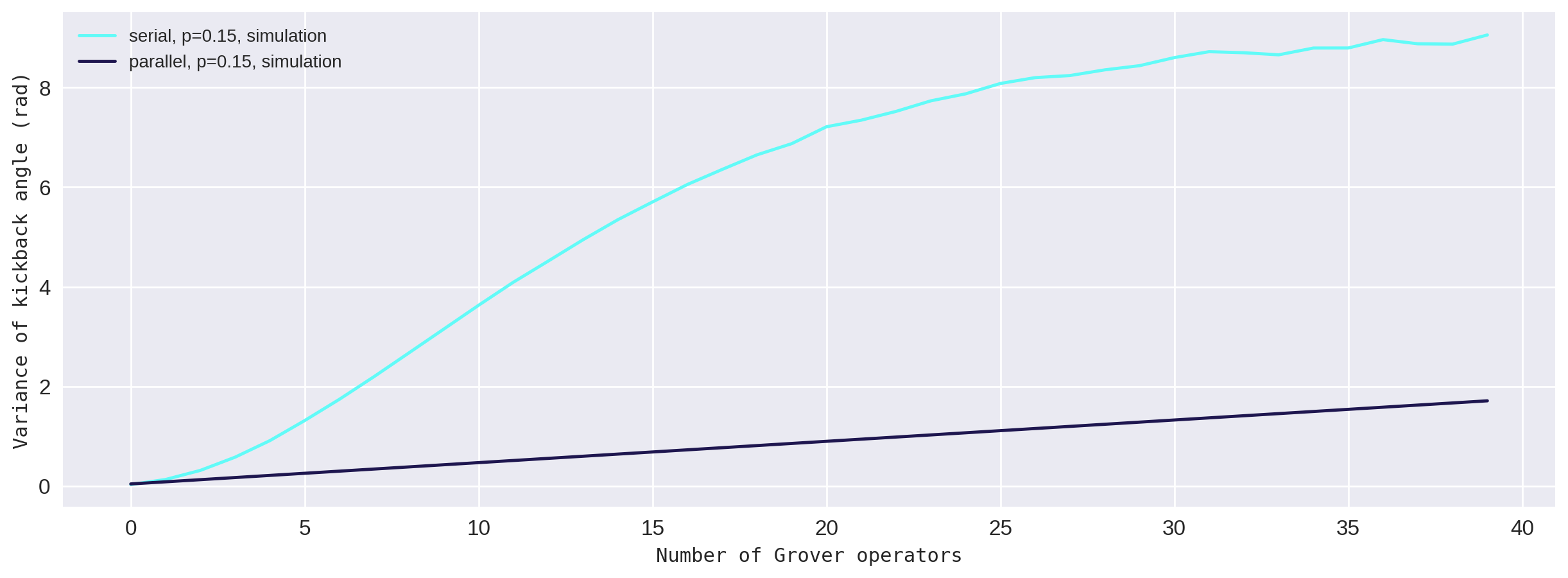}
\caption{The variance of the angle kicked back as a function of the number of Grover operators. We see that the parallel setup shows a significantly lower variance than the serial setup.}
\label{plotkickbackvar}
\end{figure}

This shows that the key feature of the case without errors -- the linearity of the kickback in the number of operators -- is preserved in the parallel setup, unlike in the serial setup as discussed in section \ref{errorsinamplitudeestimation}. However, the slope of the line is changed by the factor $1-p$.\\
When working with such circuits in practice, we have to decide how many measurements we want to perform. This is also relevant for the question if such a procedure can achieve quantum advantage in the presence of hardware errors. The fact that the number of successful kickbacks is represented by a binomial distribution implies the variance:
$$
{\rm Var}(2\theta k) = \theta p (1-p) N
$$
So we see a linear increase of the variance of the phase with the number of Grover operators. In practice, this can be good enough to achieve quantum advantage, but it will reduce the quadratic advantage of the quantum algorithm over the classical Monte Carlo solution. This is illustrated in figure \ref{plotkickbackvar}.

After analyzing the angles of the kickbacks we look at the behavior of the measurements in low depth circuits as in figure \ref{LDPhaserCircuit}. This corresponds to calculating the expectation value of ${\rm cos}(X)$ where $X$ is a binomially distributed random variable.\\

First, note that the binomial distribution with $N$ draws and a probability of $1-p$ for good events, which is the successful execution of the operator in our case, can be approximated by a normal distribution with mean value $\mu$ and variance $\sigma^2$ given by
$$
\mu = N (1-p) \quad {\rm and} \quad
\sigma^2 = Np(1-p).
$$
The accuracy of this approximation is good enough even for small examples with just a few operators $G$.

Next, we rewrite our random variable $X$ in terms of the mean $\mu$ and a new random variable $Z$ which is distributed according to a standard normal distribution:
\begin{equation}
X = \sigma Z + \mu
\end{equation}
Then the calculation of the expectation value is relatively straightforward:
\begin{eqnarray}
\mathbb{E}({\rm cos}(2 \theta X)) &=& \mathbb{E} \left(\frac{e^{i 2 \theta X }+e^{-i 2 \theta X}}{2}\right)\\
&=& \frac{1}{2}\left(e^{i 2 \theta \mu} \mathbb{E}\left(e^{i 2\theta \sigma Z}\right) + e^{-i 2 \theta \mu} \mathbb{E}\left(e^{-i 2 \theta \sigma Z}\right)\right)\\
&=& e^{\frac{-\sigma^2 2 \theta}{2}} \frac{e^{i 2 \theta \mu} + e^{-i 2 \theta \mu}}{2}\\
&=& e^{\frac{-\sigma^2 2 \theta}{2}} {\rm cos}(\mu 2 \theta)
\end{eqnarray}
This means that for the parallel phase estimation the probability $P_1$ for measuring the result~$1$ in a circuit as in figure~\ref{LDPhaserCircuit} is given by an exponentially dampened oscillation:
\begin{equation}
\label{phaserexpected}
P_1 = \frac{1 - {\rm cos}(X)}{2} = \frac{1}{2}-\frac{ e^{\frac{-\sigma^2 2 \theta}{2}}{\rm cos}(\mu 2 \theta)}{2}.
\end{equation}
At first, this sounds discouraging as this means that the signal we want to measure has an exponential dampening factor. However, in practice, the asymptotic behavior is not what matters but rather the behavior in the parameter range that we actually analyze.\\

We can simulate the behavior of such a circuit in the presence of noise. We use the circuit from section~\ref{StandardLowDepth} and our noise model is as follows:  For each Grover operator $G$ we add an additional $X$ gate on a random qubit of the Grover register with probability $p$ before $G$.
For each initialization EP we add an additional $X$ gate with probability $p/2$ before the operation on a randomly chosen qubit. This reflects the complexity of the approximate EP gate that we discuss in section~\ref{sectionEP}.
We can then compare the theoretical result of equation~\ref{phaserexpected} with the results from the simulations and with the corresponding results from the serial case of figure~\ref{plotLDcomparison2}. We find the following:
\begin{figure}[H]
\centering
\includegraphics[width=0.8\textwidth]{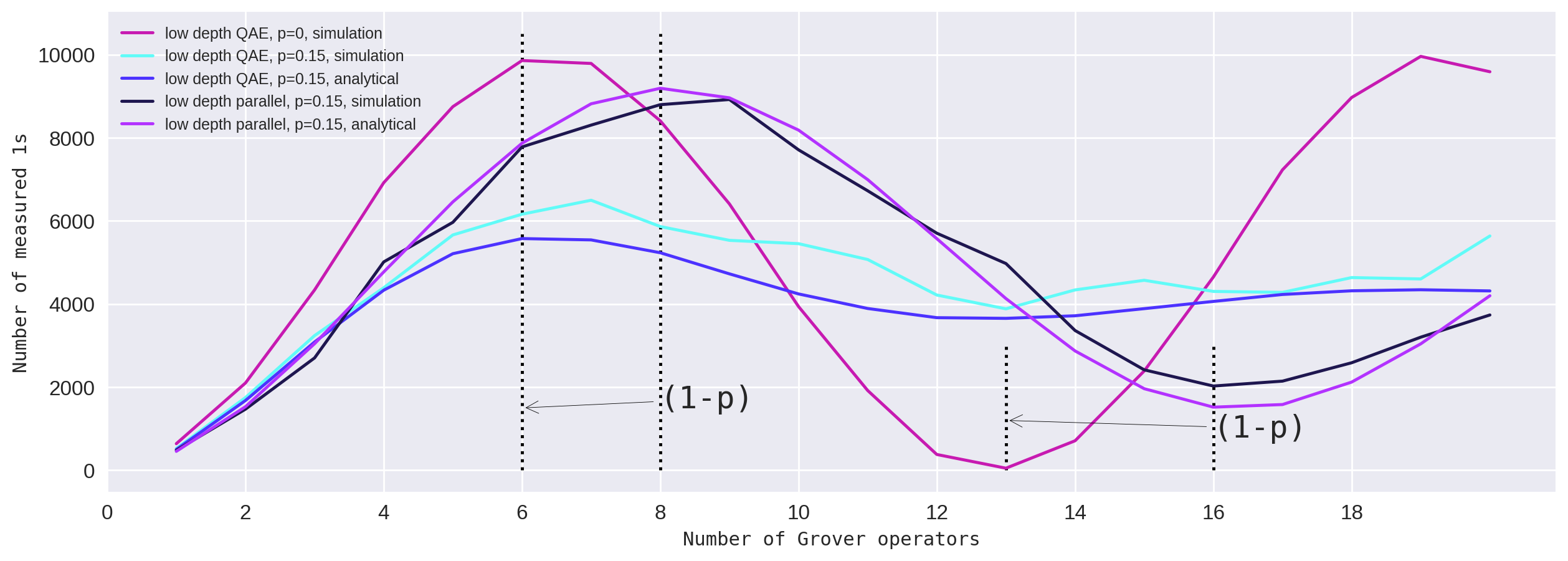}
\caption{The measurement results we find for the circuits in figure~\ref{PAEWPE} and~\ref{LDPhaserCircuit} for an increasing number of Grover operators with 10000 shots per experiment. The circuit simulations with errors were done 100 times, each with randomly drawn errors. The effect of errors can be clearly seen: The output of the parallel circuit is stretched along the x-axis by a factor equal to the error rate and it is dampened, as described by equation~\ref{phaserexpected}. The output of the serial low depth QAE circuit cannot be expressed in closed form as it contains a cosine integral. Nevertheless, it can be calculated numerically.
Note that in this example, the effective error for the parallel execution is $1-(1-p)\cdot (1-p/2)=0.21375$, since we add an error for a Grover operator with the probability $p$ and for a state preparation with the probability $p/2$.}
\label{plotLDcomparison3}
\end{figure}

The information that amplitude estimation methods extract from these measurement results is the frequency (or wavelength) of the oscillation. The accuracy of this information is of key importance. For the parallel version of amplitude estimation the signal gets stretched by a factor of $1-p$. Since it is possible to estimate $p$ in advance by building a circuit of the same complexity as the operator and measuring its fidelity, this factor can be estimated. Therefore, we can extract the correct information from the signal.\\

For extracting the information about the frequency of the oscillation we want to perform as few measurements as possible. It is particularly undesirable if the number of required measurements grows with the number of Grover operators. The number of necessary measurements is related to the amplitude of the signal: The bigger the amplitude, the fewer measurements are necessary to derive the information.\\

It can be seen from figure~\ref{plotLDcomparison3} that the parallel version provides a better amplitude
than the serial version. Therefore, the parallel version requires fewer shots than the serial standard version of QAE. Furthermore, formula~\ref{kapprox} shows  that the signal disappears quickly in the case of a serial QAE. As a rule of thumb, the formula shows that the number of operators that is reliable is proportional to the inverse error rate $1/p$.
On the other hand, the parallel version has an exponential dampening of the signal, but the signal remains usable even for a large number of operators. A further disadvantage of the serial QAE can be seen for the non low depth version as in figure~\ref{PE}: The expectation value per operator is different on each qubit that gathers kickbacks. The parallel setup does not have this drawback.

\subsection{No-Go Theorems for Parallel Approximate Amplitude Estimation}
The parallelization of QAE for approximate counting has been discussed in the literature and there is a no-go theorem that states that the parallelization only leads to an improvement that is as good as a classical parallelization~\cite{Burchard}.
For the proof of this theorem an operator $O_x$ on $n$ qubits is considered and $x$ is a vector of length $2^n$ that shows for input $|i\rangle=|i_1\ldots i_n\rangle$ the value $0$ or $1$ for the operator. The task is to find an approximation of the number of ones, when the operator applies  the phase $+1$ to inputs $|i\rangle$ with $O_x(i)=0$ and the phase $-1$ for inputs $|i\rangle$ with $O_x(i)=1$. The standard QAE can be used to solve this problem and the bound of~\cite{Burchard} applies.

For our parallel version of QAE, we assume that besides the operator $O_x$ we additionally have a gate EP that generates one of the eigenstates
$$
|\Psi_\pm\rangle=\frac{1}{\sqrt{2}} \left(|{\tilde \Psi}_1\rangle \pm
i|{\tilde \Psi}_0\rangle \right)
$$
of the Grover operator, where $|{\tilde \Psi}_i\rangle$ with $\langle {\tilde \Psi}_i | {\tilde \Psi}_i\rangle =1$ are normalized versions of the vectors $|\Psi_0\rangle$ and $|\Psi_1\rangle$
that are superpositions of all bad and good states, respectively.\\
If we have one of these eigenstates and we measure  whether a state is good or bad then we obtain a good state with 50\% probability and this is independent from the size $2^n$ of the search space. Therefore, for a small number of good states, this would allow us to trivially solve the counting problem by sampling elements from the set of good states. This shows that we cannot apply the no-go theorem for the approximate counting problem if we assume that a single eigenstate of the Grover operator can be constructed or approximated.

\section{Eigenstate Preparation}\label{sectionEP}
The standard method for QAE for an operator $M$ starts with the application of $M$ on the quantum register on which the corresponding Grover operators act. The operator $M$ generates a superposition of the two relevant eigenvectors $|\Psi_+\rangle$ and $|\Psi_-\rangle$ of the Grover operator~\cite{QAE}. Since there is only one quantum register with this superposition, the standard QAE calculates a binary representation for each corresponding eigenvalue in superposition. The final measurement leads to one of the two values and both lead to the same final result. However, if we initialize different quantum registers with the operator $M$, we have a superpositions on each part of the system and this destroys the result. Therefore, we have to generate a single eigenstate $|\lambda\rangle$ of the Grover operator on each register.

In some special cases, the preparation of a single eigenstate can be achieved efficiently and exactly. The simplest such example is a Grover operator on one qubit as discussed in section~\ref{errorsin2d}.
In some cases, when an exact preparation of the eigenstate is not possible then  we can  construct an approximation. The construction and the application to an example is shown in the following.

\subsection{Eigenstate Approximation}\label{secEP}
We show how an approximation of a single eigenstate $|\lambda\rangle$ can be constructed under the following assumptions:
\begin{itemize}
\item There is only one good state, i.e. the operator $M$ generates the state $M|0\rangle=|\Psi_0\rangle + |\Psi_1\rangle$ and the good state $|\Psi_1\rangle$ is a single state in the computational base.
\item The good state $|\Psi_1\rangle$ has a small probability, i.e.
$a=|\langle \Psi_1 | \Psi_1 \rangle |^2$ is small.
\end{itemize}
We also assume that there is an efficient circuit for applying the phase $-1$ to $|\Psi_1\rangle$ and the phase
$+1$ to $|\Psi_0\rangle$. We denote this operation by $P$.
Furthermore, we assume that the normalized state $|{\tilde \Psi}_1\rangle$, i.e. the same basis state but with $|\langle {\tilde \Psi}_1 | {\tilde \Psi}_1\rangle|=1$, can be built efficiently with a quantum circuit.

To simplify notation, we first consider the construction in a more general setting: We consider matrices $V$ and $W$ that turn $|0\rangle$ to $|v\rangle$ and $|w\rangle$, respectively. We first assume that $|v\rangle$ and $|w\rangle$ are orthogonal and that we can apply a phase operation $P$ with $P|v\rangle=|v\rangle$ and $P|w\rangle=-|w\rangle$. Then we can create the superposition $\frac{1}{\sqrt{2}}(|v\rangle+|w\rangle)$ with the following circuit:

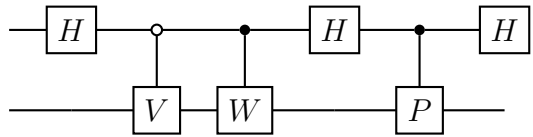
\begin{figure}[H]
\centering
\begin{quantikz}
& \gate{H} & \octrl{1} & \ctrl{1} & \gate{H} & \ctrl{1} & \gate{H} & \\
& \qw      & \gate{V}  & \gate{W} & \qw      & \gate{P} & \qw      &
\end{quantikz}
\caption{Simplified circuit scheme for the approximation of an eigenvector of a Grover operator.}
\label{UandV}
\end{figure}

The first Hadamard gate and the two controlled $V$ and $W$ operations generate
$$
\frac{1}{\sqrt{2}}\left( |0\rangle |v\rangle+|1\rangle|w\rangle\right).
$$
The second Hadamard gate leads to
$$
\frac{1}{2}\left(|0\rangle(|v\rangle+|w\rangle)+|1\rangle(|v\rangle-|w\rangle)\right)
$$
and after the controlled $P$ operation we obtain
$$\frac{1}{2}\left(|0\rangle(|v\rangle+|w\rangle)+|1\rangle(|v\rangle+|w\rangle)\right)=\frac{1}{2}\left(|0\rangle+|1\rangle\right)\cdot\left(|v\rangle+|w\rangle\right)
$$
The final Hadamard transform then leads to the state $\frac{1}{\sqrt{2}}\left(|v\rangle+|w\rangle\right)$ when we ignore the first qubit that is restored to $|0\rangle$. Note, that when we start with the state
$|1\rangle |0\rangle$ instead of $|0\rangle|0\rangle$ then we obtain the state $\frac{1}{\sqrt{2}}\left(|v\rangle-|w\rangle\right)$ at the end.

In our case, we have $|v\rangle=|\Psi_0\rangle+|\Psi_1\rangle$ and $|w\rangle=|{\tilde \Psi}_1\rangle$ with $V=M$ and an operator $W$ with $W|0\rangle=|{\tilde \Psi}_1\rangle$. Since $|{\tilde \Psi}_1\rangle$ is a single basis vector, the matrix $W$ can be constructed easily. The states $|v\rangle$ and $|w\rangle$ are not orthogonal and this means that we cannot find a unitary $P$ with $P|v\rangle=|v\rangle$ and $P|w\rangle=-|w\rangle$ as required above. However, the operator $P$ from the beginning of this section is an approximation for it because $|v\rangle$ and $|w\rangle$ are almost orthogonal since $|\Psi_1\rangle$ has
a very small magnitude by assumption. Taking this into account, the result at the end of the circuit is
\begin{equation}\label{eqApprox}
\frac{1}{\sqrt{2}}|0\rangle\left( |\Psi_0\rangle+|{\tilde \Psi}_1\rangle)\right) + \frac{1}{\sqrt{2}}|1\rangle|\Psi_1\rangle
\end{equation}

\begin{figure}[H]
\centering
\includegraphics[width=0.8\textwidth]{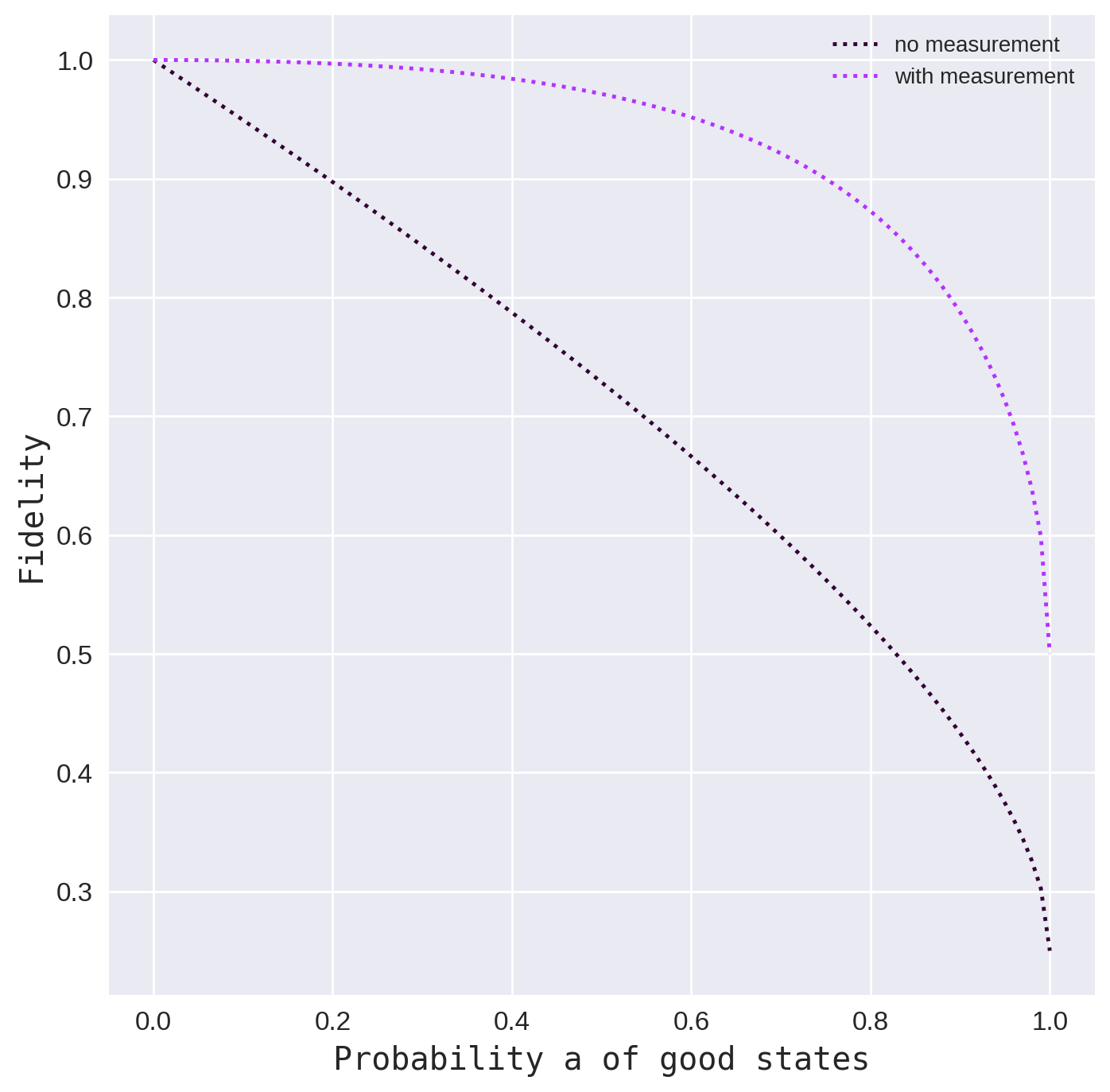}

\caption{Fidelities of both approximation methods. The fidelity of the method with measurement is much higher than the fidelity of the method without measurement. Note that the measurement-based method only produces a valid state with probability $1-a$.}
\label{plotFidelities}
\end{figure}

For $|\Psi_1\rangle$ with small amplitude we have $|{\tilde \Psi}_0\rangle \approx |\Psi_0\rangle$ and $|\Psi_1\rangle \approx 0$ and therefore we have an approximation of the desired state
$$
\frac{1}{\sqrt{2}}|0\rangle\left( |{\tilde \Psi}_0\rangle + |{\tilde \Psi}_1\rangle \right)
$$
The scalar product of this state, which is the exact one  we want to construct, and the state in equation~\ref{eqApprox} can be calculated to be
$$
\frac{\sqrt{1-a}+1}{2}
$$
and the fidelity depending on the parameter $a$ is shown in figure~\ref{plotFidelities}, where it is the curve labeled 'no measurement'. With an additional operator that applies the phase $i$ to $|{\tilde \Psi}_0\rangle$ we obtain an approximation of one of the eigenvectors of the Grover operators as defined in~\cite{QAE}.

The quality of the approximation can be improved by performing a measurement on the first qubit after the construction of the state in equation~\ref{eqApprox}. We obtain the unwanted result $1$ with the probability $a$ which is small by our assumptions. If this happens, we do not consider the generated state any further. If we obtain the result $0$ then the resulting state is
$$
\frac{1}{\sqrt{2-a}}\left(|\Psi_0\rangle+|{\tilde \Psi}_1\rangle \right)
$$
when we neglect the measured qubit. We can rewrite this as
$$
\frac{1}{\sqrt{2-a}}\left(|{\tilde \Psi}_0\rangle+|{\tilde \Psi}_1\rangle \right)
+\frac{1}{\sqrt{2-a}}\left(|\Psi_0\rangle-|{\tilde \Psi}_0\rangle\right)
$$
where $|{\tilde \Psi}_0\rangle \approx |\Psi_0\rangle$ shows that the second part is small.

The scalar product of the desired state $\frac{1}{\sqrt{2}}\left(|{\tilde \Psi}_0\rangle+|{\tilde \Psi}_1\rangle\right)$ and this state is
$$
\frac{\sqrt{1-a}+1}{\sqrt{2(2-a)}}
$$
The fidelity of this approximation depending on the parameter $a$ is shown in figure~\ref{plotFidelities}, where it is the curve labeled 'with measurement'

\subsection{Example for Standard QAE with Eigenstate Approximation}\label{sectionApprox}
To illustrate the construction of the approximation of an eigenvector of a Grover operator we consider an example with a standard QAE that uses the following unitary $M$:
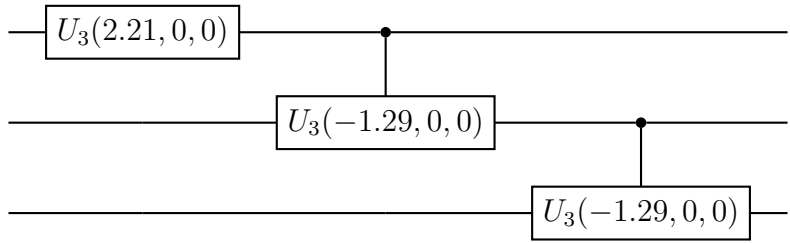
\begin{figure}[H]
\centering
\begin{quantikz}
\qw&\gate{U_3(2.21,0,0)} &\ctrl{1}   &\qw        &\qw \\
\qw&\qw        &\gate{U_3(-1.29,0,0)} &\ctrl{1}   &\qw \\
\qw&\qw        &\qw        &\gate{U_3(-1.29,0,0)} &\qw
\end{quantikz}

\caption{An example for a unitary $M$ that is used to construct a Grover operator. We use the notation $U_3$ from Qiskit~\cite{qiskit}.}

\label{approxExampleA}
\end{figure}

We obtain the following distribution when the three qubits are measured after the application of $M$:

\begin{figure}[H]
\centering
\includegraphics[width=0.6\textwidth]{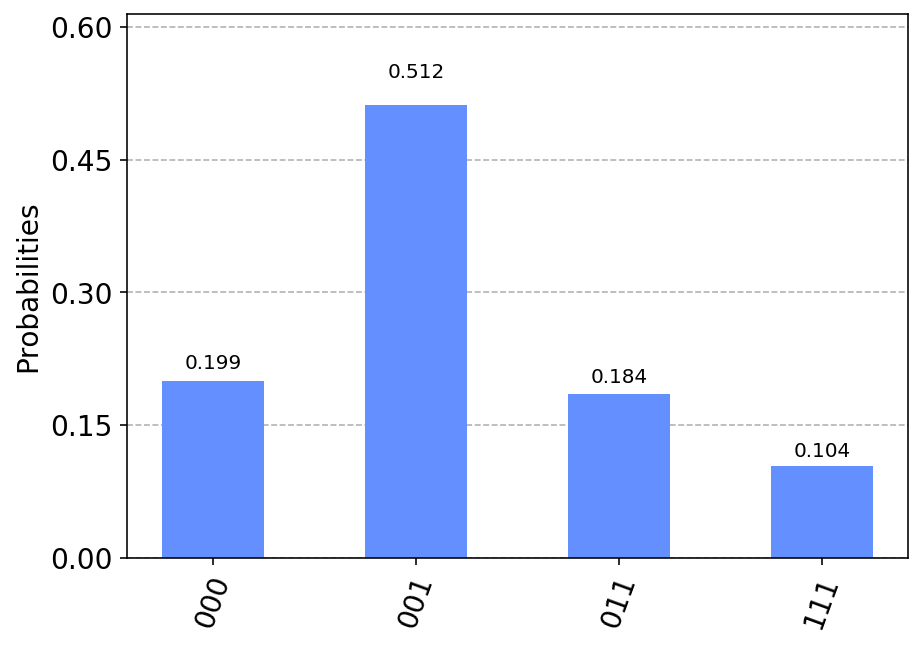}
\caption{The histogram of a measurement in the standard basis after running the unitary $M$ from figure~\ref{approxExampleA}.}
\label{histoApproximation}
\end{figure}

The good state we are looking for is $|111\rangle$ and it has a probability of $0.104$. A possible circuit for the Grover operator for the unitary $M$ and the good state $|111\rangle$ looks like follows:

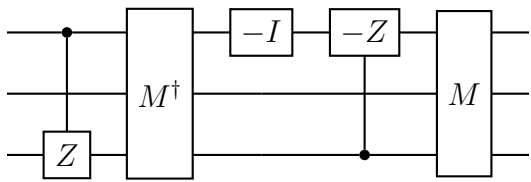
\begin{figure}[H]
\centering
\begin{quantikz}
\qw & \ctrl{2} & \gate[wires=3]{M^\dagger} & \gate{-I} & \gate{-Z} & \gate[wires=3]{M} & \qw \\
\qw & \qw      &                           & \qw       & \qw      &                   & \qw \\
\qw & \gate{Z} &                           & \qw       & \ctrl{-2}&                   & \qw
\end{quantikz}
\caption{Circuit that implements the Grover operator for $M$ and the good state $|111\rangle$.}
\label{approxExampleA2}
\end{figure}

The two relevant eigenvalues of this operator are $\lambda=0.7926\pm 0.6097i$, all other eigenvalues are $-1$. The arguments of these complex numbers are ${\rm arg}(\lambda)=\pm 0.656$. We obtain the probability ${\rm sin}({\rm arg}(\lambda)/2)=0.104$ of the good state for both eigenvalues~\cite{QAE}.

Following the construction in the previous section, a circuit for approximating one eigenvector of this Grover operator looks like follows:

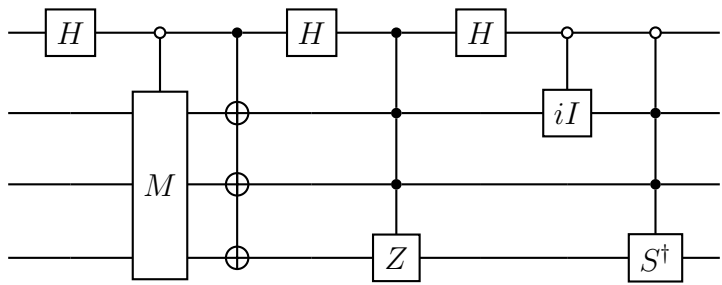
\begin{figure}[H]
\centering
\begin{quantikz}
\qw & \gate{H}          & \octrl{1}         & \ctrl{3}  & \gate{H} & \ctrl{1} & \gate{H} & \octrl{1} & \octrl{1}       & \qw\\
\qw & \qw               & \gate[wires=3]{M} & \targ{}   & \qw      & \ctrl{1} & \qw      & \gate{iI}  & \ctrl{1}        & \qw \\
\qw & \qw               &                   & \targ{}   & \qw      & \ctrl{1} & \qw      & \qw       & \ctrl{1}        & \qw \\
\qw & \qw               &                   & \targ{}   & \qw      & \gate{Z} & \qw      & \qw       & \gate{S^\dagger}& \qw
\end{quantikz}
\caption{Circuit for the approximation of an eigenstate of the Grover operator where $S^\dagger={\rm diag}(1,-i)$. By using the complex conjugated versions of $iI$ and $S^\dagger$ we obtain the other eigenvector of the Grover operator.}
\label{EPEsimple2}
\end{figure}

In this case, calculations show that the fidelity of the real eigenvector and the approximation constructed by this circuit is $0.947$. The following circuit is used to estimate the probability $0.104$ of the good state with a precision of five bits by
standard QAE. Note that the circuit for the approximation of the eigenvector uses an additional qubit that is not shown in the circuit.

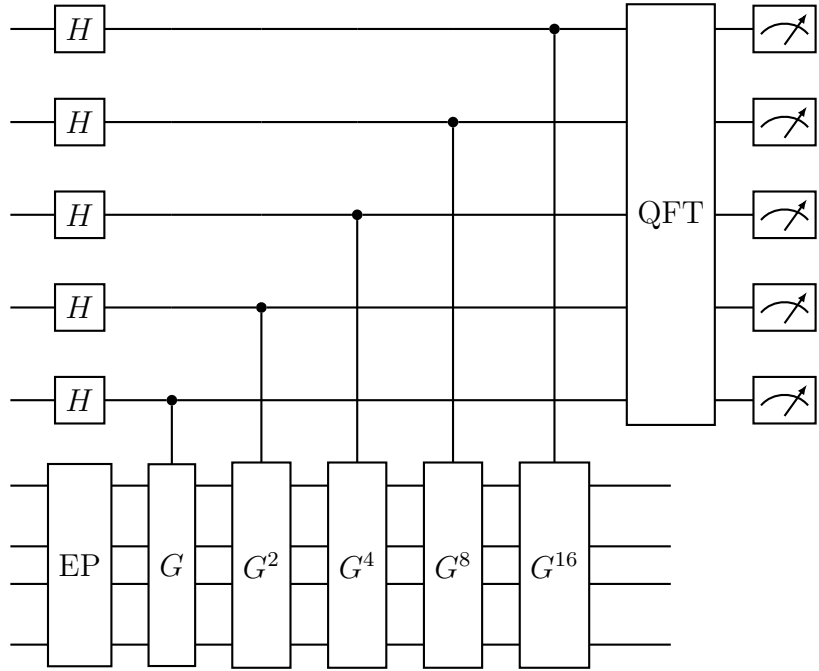
\begin{figure}[H]
\centering
\begin{quantikz}
\qw & \gate{H}               & \qw                   & \qw                   & \qw                   & \qw                   & \ctrl{5}               & \gate[wires=5]{\rm QFT}&\meter{} \\
\qw & \gate{H}               & \qw                   & \qw                   & \qw                   & \ctrl{4}              & \qw                    &                        &\meter{} \\
\qw & \gate{H}               & \qw                   & \qw                   & \ctrl{3}              & \qw                   & \qw                    &                        &\meter{} \\
\qw & \gate{H}               & \qw                   & \ctrl{2}              & \qw                   & \qw                   & \qw                    &                        &\meter{} \\
\qw & \gate{H}               & \ctrl{1}              & \qw                   & \qw                   & \qw                   & \qw                    &                        &\meter{} \\
\qw & \gate[wires=4]{\rm EP} & \gate[wires=4]{G}     & \gate[wires=4]{G^{ 2}}& \gate[wires=4]{G^{ 4}}& \gate[wires=4]{G^{ 8}}& \gate[wires=4]{G^{16}} & \qw                    &\\
\qw &                        &                       &                       &                       &                       &                        & \qw                    &\\
\qw &                        &                       &                       &                       &                       &                        & \qw                    &\\
\qw &                        &                       &                       &                       &                       &                        & \qw                    &
\end{quantikz}
\caption{Circuit for the QAE with five bits precision.}
\label{EPEsimple3}
\end{figure}

As the standard QAE is using only unitary operations we have the same fidelity before the measurement at the end. Therefore, we expect to obtain the same result as with the exact eigenvector with a high probability. The following histogram shows the measurement results for three different unitaries EP for the initialization: (1) ${\rm EP}=M$ leading to a superposition of two eigenvectors, (2) one exact eigenvector, (3) the approximation of the eigenvector. 

\begin{figure}[H]
\centering
\includegraphics[width=1.0\textwidth]{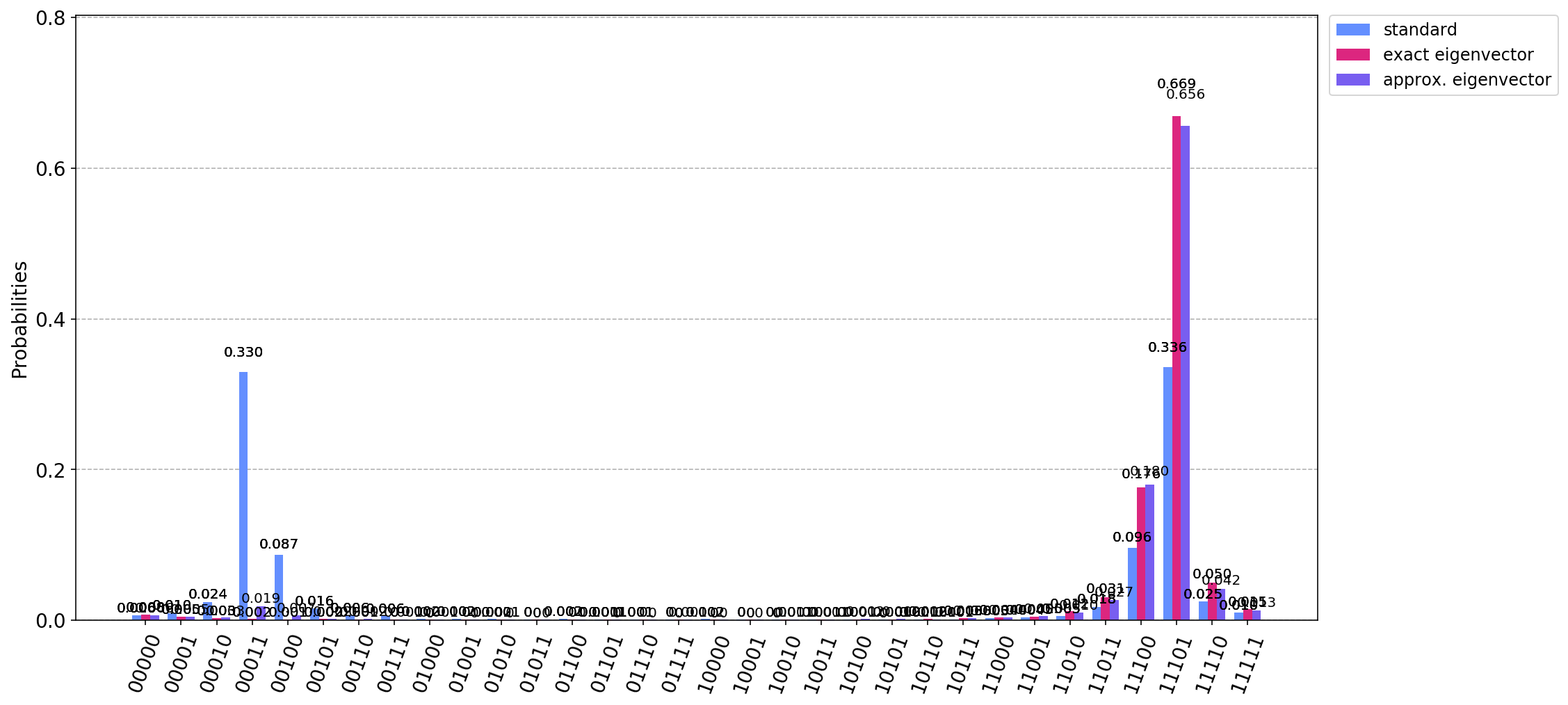}
\caption{Result histograms for QAE with three different unitaries as gate EP.}
\label{histoApproximation2}
\end{figure}

When we map the binary measurement results to the corresponding probability values then we obtain the following distribution of results.

\begin{figure}[H]
\centering
\includegraphics[width=1.0\textwidth]{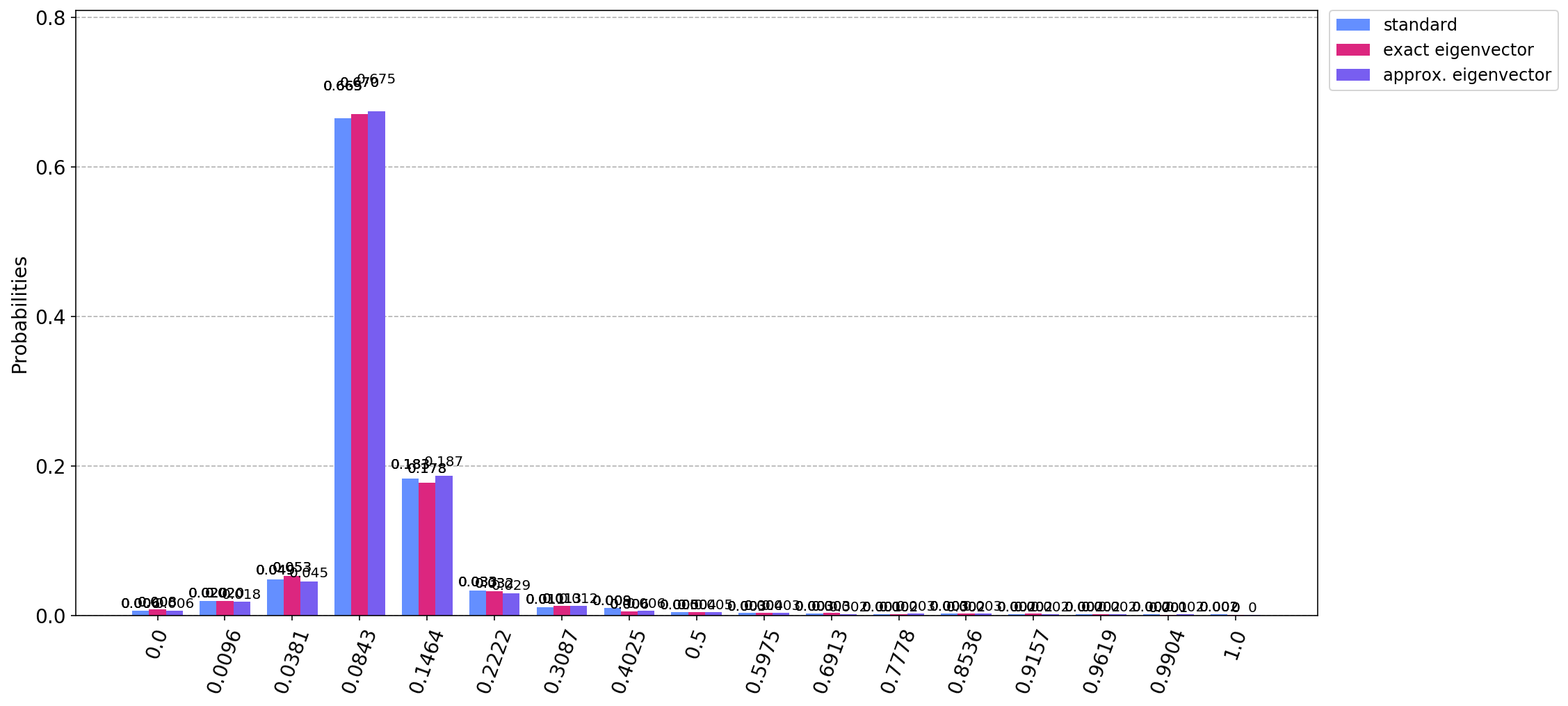}
\caption{Result histogram showing standard QAE (light blue) vs. QAE with one exact eigenvector (red) vs. QAE with
approximation of one eigenvector (purple). Here we consider the probability values that correspond to the binary measurement results.}
\label{histoApproximation3}
\end{figure}

Note that with the resolution of five bits, the result $0.084$ is closest to the real solution and that the binary solutions $00011$ and $11101$ are mapped to this value. Therefore, with a probability of $67\%$ we obtain the result that is closest to the real value.

\section{Algorithm Specific Error Correction}\label{sectionErrorCorrection}
In the examples we have discussed in section~\ref{secErrorGrover}, the kickback after an error on the Grover register is $-1$. Therefore, an odd number of erroneous $-1$ kickbacks results in a $-1$ kickback whereas an even number results in a $+1$ kickback. The big difference between the two cases makes the estimation of the error very hard, because we do not know the number of errors. To tackle this problem, we show a way to extend the parallel version of QAE with an algorithm-specific error correction that checks if a Grover operator was correctly executed and corrects the wrong kickback.

While the idea of detecting and correcting errors will work for errors on the Grover register before the
execution of a Grover operator, it does not work if the error is inside a Grover operator 
because such errors do typically not lead to $-1$ kickbacks, see the discussion at the end of 
section~\ref{errorsingrovers}. Although we do not gain information about the erroneous kickback in
this case with our construction, we still can use the information that there was an error by flagging the result of a measurement as faulty and discarding it.

\subsection{Construction with Inverse Eigenstate Preparation}
The main idea for an error correction is the following: If we can efficiently construct a circuit EP  which prepares a suitable eigenstate $|\lambda\rangle$ of the Grover operator, then we can also invert it. This means that in the circuit in figure~\ref{PPE} we can simply follow each Grover operator with the inverse ${\rm EP}^\dagger$ of the eigenstate preparation. As long as there was no error, this will lead to the state $|0\ldots0\rangle$ of the register on which the Grover operator acts. Any other state signals that an error occurred and we could disregard the result or count the number of errors and try to fix the error by using this information.
For the cases where an error leads to a $-1$ kickback we could also execute a $Z$ gate on the qubit which collects the kickbacks. This way, we retain a small error because we get no kickback instead of the correct kickback, which is small by the assumptions in section~\ref{secErrorGrover}, but we correct the big phase error $-1$.

It is possible to add another layer of error correction. For this, a state that indicates an error controls
the $-1$ kickback and additionally another kickback from a Grover operator on another register. This way, not only
the $-1$ kickback is corrected but also the missing kickback is fixed. It is possible to add more layers of error correction. However, the extra qubits and complexity will usually not be worth the overhead, especially when one can estimate the error rate and compensate for it in post processing. 

\begin{figure}[H]
\centering
\begin{quantikz}
& \gate{H}              & \ctrl{1}          & \gate{Z}                     & \gate{Z}  & \ctrl{4}          & \gate{Z}                      & \gate{Z}  & \gate{H} &\qw \\
& \gate[wires=3]{\rm EP}& \gate[wires=3]{G} & \gate[wires=3]{{\rm EP}^\dagger}& \octrl{-1}& \qw               & \qw                           & \qw       & \qw      &\qw \\
&                       &                   &                              & \octrl{-1}& \qw               & \qw                           & \qw       & \qw      &\qw \\
&                       &                   &                              & \octrl{-1}& \qw               & \qw                           & \qw       & \qw      &\qw \\
& \gate[wires=3]{\rm EP}& \qw               & \qw                          & \qw       & \gate[wires=3]{G} & \gate[wires=3]{{\rm EP}^\dagger} & \octrl{-4}& \qw      &\qw \\
&                       & \qw               & \qw                          & \qw       &                   &                               & \octrl{-1}& \qw      &\qw \\
&                       & \qw               & \qw                          & \qw       &                   &                               & \octrl{-1}& \qw      &\qw
\end{quantikz}
\caption{In this example for a circuit with corrections of erroneous $-1$ kickbacks, the Grover register has 3 qubits and there are two operators kicking back onto a single qubit. The conditioned $Z$ gates reverse the unconditioned $Z$ gates for the state $|000\rangle$ that indicates that no error happened.
This can be seen as an example of a low depth QAE circuit, but if the second H gate is left out, it can act as a part of a full QAE circuit.}
\label{ECbyEP}
\end{figure}
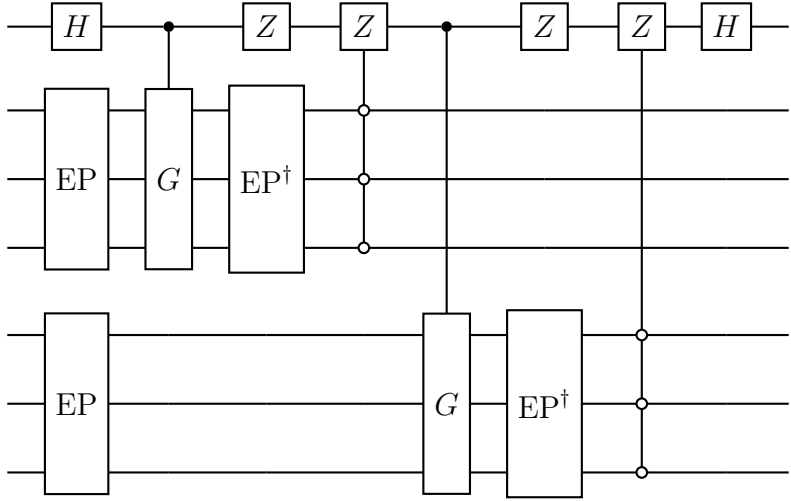

And example for a circuit with the correction of erroneous $-1$ kickbacks is shown in figure~\ref{ECbyEP} for a parallel low depth QAE  with two phase kickbacks.

A problem with this method is the potentially high complexity of the EP gate in some cases. If we use the approximation circuit of section~\ref{sectionEP} then each EP gate includes an operator $M$ that is controlled by one qubit. Therefore, the complexity of the correction with the gate ${\rm EP}^\dagger$ is roughly the same as the complexity of the controlled Grover operator.
Therefore, if we assume that the controlled Grover operator, which has the two controlled operations $M$ and $M^\dagger$, is prone to errors, then we also have to assume that EP and ${\rm EP}^\dagger$ are also prone to errors. This reduces the efficiency of the error correction method significantly.

\subsection{Construction with Intermediate Measurements}\label{secECmeasure}
To avoid the high complexity of the controlled ${\rm EP}^\dagger$ gate of the construction in the previous section, we can replace it with a gate that is controlled by a measurement result. This is useful if a hardware implementation of a quantum computer allows measurements during a calculation and also operations that are controlled by measurement results.

For this construction, we consider the approximation circuit of figure~\ref{UandV} and remove the first two gates from the left. We call this circuit ${\rm EP}_2$. The circuit looks like follows:
\begin{figure}[H]
\centering
\begin{quantikz}
&  \ctrl{1} & \gate{H} & \ctrl{1} & \gate{H} & \qw\\
&  \gate{W} & \qw      & \gate{P} & \qw      & \qw
\end{quantikz}
\caption{Circuit schema of ${\rm EP}_2$, which is a subpart of the approximation circuit in figure~\ref{UandV}.}
\label{UandV2}
\end{figure}
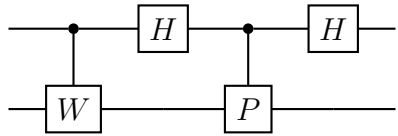

Then the state ${\rm EP}_2^\dagger \cdot {\rm EP} \cdot|0\rangle$ is 
$$
\frac{1}{\sqrt{2}}|0\rangle  |\mu\rangle + \frac{1}{\sqrt{2}}|1\rangle |0\rangle \quad {\rm with}\quad
|\mu\rangle = M|0\rangle.
$$
Therefore, if we measure the first qubit and obtain the result $1$ then we know that the second part is in state $|0\rangle$. The same can be achieved for the result $0$ if we apply $M^\dagger$ after the measurement.

The circuit for a single kick-back with the measurement-based error correction looks like follows:
\begin{figure}[H]
\centering
\begin{quantikz}
\qw                      & \qw                       & \qw                    & \ctrl{2}                       & \qw                                & \qw              &\gate{Z} &\gate{Z}   &\qw\\
\qw                      & \gate[wires=2]{|0\rangle} & \gate[wires=2]{\rm EP} & \qw                            & \gate[wires=2]{{\rm EP}_2^\dagger} & \meter{0}\vcw{1} &         &           &   \\
\qw                      & \qwbundle{q}              &  \qw                   & \gate{G}                       & \qw                                & \gate{M^\dagger} &\qw      &\octrl{-2} &\qw
\end{quantikz}
\caption{The circuit for a single kickback of an eigenvector of the Grover operator $G$ with the measurement-based error correction. The qubit in the middle is the ancilla qubit used for the approximation circuit of figure~\ref{UandV}. The last controlled operation is executed if the $q$ qubits on the bottom are in the state $|0\ldots0\rangle$.}
\label{ecmeasurement}
\end{figure}
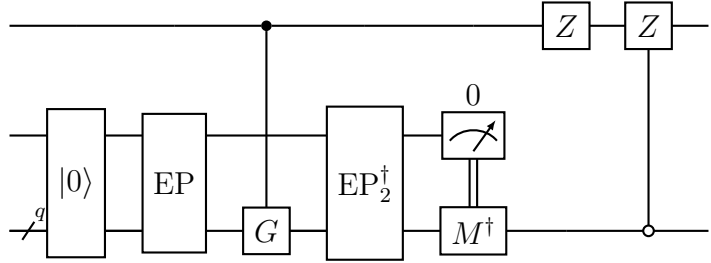
This circuit uses a classically controlled gate $M$ instead of a controlled $M$ gate. Note that the quantum controlled version of $M$ has a much higher complexity than the uncontrolled version when using standard decomposition techniques for controlled operators as described in~\cite{barenco} if only single- and two-qubit basis gates are available.

\section{Practical Application -- an Example}\label{sectionExample}
After presenting parallel versions of amplitude estimation, eigenvector approximations and methods to correct errors we apply these techniques to a small example. We choose the parallelization as described in section~\ref{secReinit}, which is based on reinitialization.
The problem we solve is finding the probability of the worst case scenario in a small scale business risk model described in section 2.1 of ref.~\cite{jos}. In this example, there are four events with a certain dependency structure, where the first two events are mutually exclusive. The quantum circuit which implements this model looks like this:
\begin{figure}[H]
\centering
\begin{quantikz}
\qw&\qw             &\qw        &\qw             &\qw             &\gate{U(0.431)} &\gate{U(1.430)} &\qw \\
\qw&\qw             &\qw        &\gate{U(0.643)} &\gate{U(1.671)} & \octrl{-1}     &\ctrl{-1}       &\qw \\
\qw&\qw             &\targ{}    &\octrl{-1}      &\ctrl{-1}       & \qw            &\qw             &\qw \\
\qw&\gate{U(2.214)} &\octrl{-1} &\qw             &\qw             & \qw            &\qw             &\qw
\end{quantikz}
\caption{A circuit implementation of the business risk model described in section~2.1 of ref.~\cite{jos}. Here, we use the notation $U(\theta)=U_3(\theta,0,0)$ with the notation $U_3$ from Qiskit.}
\label{modelgate2}
\end{figure}
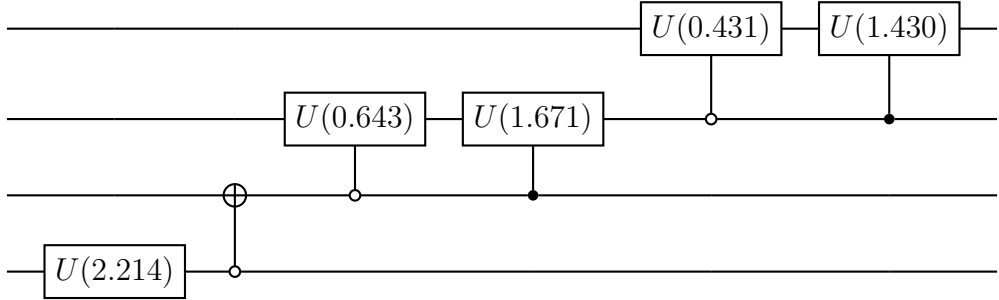

The details regarding the risk model and the construction of the circuit in figure~\ref{modelgate2} are described in~\cite{jos}. Here, we start with the gate implementing the model and show how to construct an amplitude estimation circuit to analyze it on noisy quantum hardware. Note that for relevant use cases, such a model would typically describe more than 100 events rather than the four in this example.  At this size, Monte-Carlo simulations of such a model on classical computers is still possible, but time consuming even on large parallel machines.\\

The business problem to solve here is to determine the probability of the worst case scenario, which corresponds to the output $0111$ when the four qubits in circuit~\ref{modelgate2} are measured.
We construct the Grover operator to search for the state $0111$ in the standard way. The Grover operator looks like follows:
\begin{figure}[H]
\centering
\begin{quantikz}
\qw&\ctrl{1} &\gate[wires=4]{M^\dagger}& \gate{-I} & \gate{-Z} &\gate[wires=4]{M}&\qw\\
\qw&\ctrl{1} &                         & \qw       & \octrl{-1}&                 &\qw\\
\qw&\ctrl{1} &                         & \qw       & \octrl{-1}&                 &\qw\\
\qw&\gate{-Z}&                         & \qw       & \octrl{-1}&                 &\qw
\end{quantikz}
\caption{The Grover operator as quantum circuit. The first controlled $Z$ gate applies the phase $-1$ to the state $|0111\rangle$ and the phase $+1$ to all other states.}
\label{groveroperator2}
\end{figure}
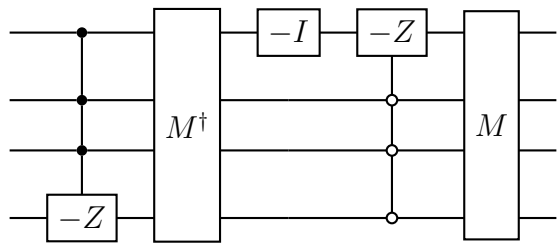

The next step is to construct the circuit EP, which prepares a single eigenstate of the Grover operator. We use our approximation technique from section~\ref{secEP} and find:
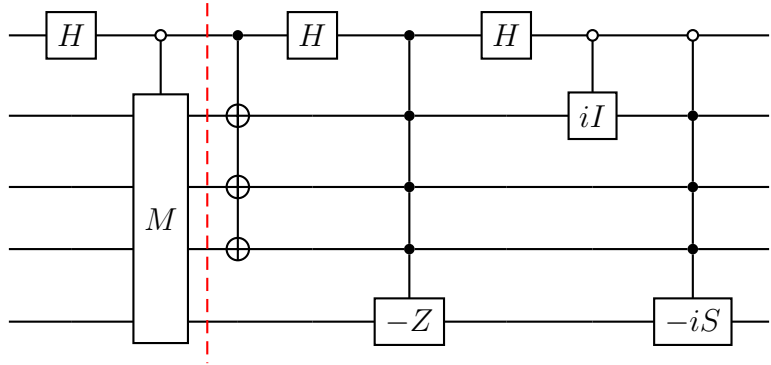
\begin{figure}[H]\label{exPrep}
\centering
\begin{quantikz}
\qw&\gate{H}&\octrl{1}\slice{}    & \ctrl{3}   & \gate{H} &\ctrl{1} &\gate{H}   & \octrl{1} &\octrl{1}  &\qw \\
\qw&\qw     &\gate[wires=4]{M} & \targ{}    & \qw      &\ctrl{1} &\qw        & \gate{iI}&\ctrl{1}  &\qw \\
\qw&\qw     &                  & \targ{}    & \qw      &\ctrl{1} &\qw        & \qw      &\ctrl{1}  &\qw \\
\qw&\qw     &                  & \targ{}    & \qw      &\ctrl{1} &\qw        & \qw      &\ctrl{1}  &\qw \\
\qw&\qw     &                  & \qw        & \qw      &\gate{-Z}&\qw        & \qw      &\gate{-iS}&\qw
\end{quantikz}
\caption{The approximate state preparation EP for the model $M$ specified in figure \ref{modelgate2}. We have $S={\rm diag}(1,i)$. 
The gate ${\rm EP}_2$ is the part on the left-hand side of the separation line.}
\end{figure}

Since the Grover operator does not satisfy the conditions for well-behaved errors as in section~\ref{secErrorGrover}, we want to apply an error correction. We choose the measurement based correction of section~\ref{secECmeasure}. The circuit for this algorithm specific error correction looks like follows:
\begin{figure}[H]
\centering
\begin{quantikz}
\qw             &\qw                    & \ctrl{2}          &\qw                              &\qw\slice{}                   & \gate{Z} &\gate{Z}    &\qw \\
\gate{|0\rangle}&\gate[wires=5]{\rm EP} & \qw               &\gate[wires=5]{\rm EP_2^\dagger} &\meter{0}\vcw{1}           & \qw      &\qw         &\qw \\
\gate{|0\rangle}&                       & \gate[wires=4]{G} &                                 &\gate[wires=4]{M^\dagger}  & \qw      &\octrl{-2}  &\qw \\
\gate{|0\rangle}&                       &                   &                                 &                           & \qw      &\octrl{-1}  &\qw \\
\gate{|0\rangle}&                       &                   &                                 &                           & \qw      &\octrl{-1}  &\qw \\
\gate{|0\rangle}&                       &                   &                                 &                           & \qw      &\octrl{-1}  &\qw 
\end{quantikz}
\caption{A circuit that produces a kickback on the qubit on the top. This circuit includes an error correction. If the measurement result is $0$ then the gate $M^\dagger$ is executed. The gates $|0\rangle$ are non-unitary reset gates and are included because we use the construction from section~\ref{secReinit}.}
\label{figureErrorMitigation}
\end{figure}
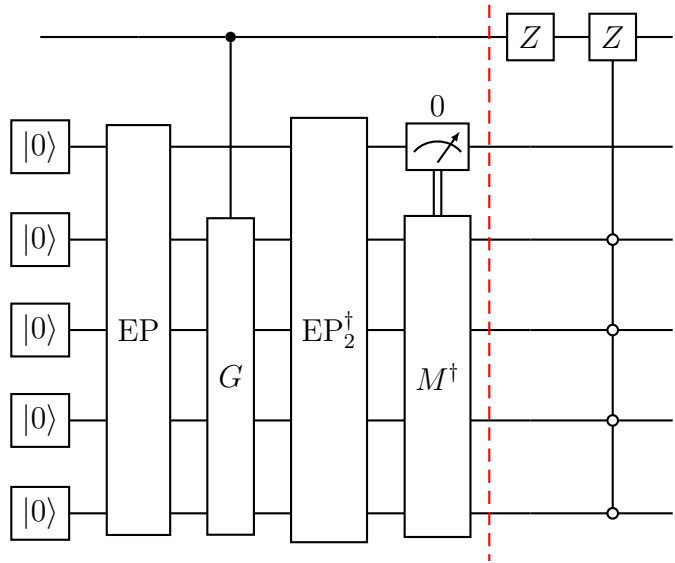

Now we have all ingredients for the parallel version of the QAE. The structure of the circuit looks like this:
\begin{figure}[H]
\centering
\begin{quantikz}
\qw         &\gate{H}       &\qw              &\qw       &\qw                   &\ctrl{2}         &\gate{Z}  &\gate{Z}              &\ctrl{2}         &\gate{Z}  &\gate{Z}               &\gate[wires=2]{\rm QFT}&\qw\\
\qw         &\gate{H}       &\ctrl{1}         &\gate{Z}  &\gate{Z}              &\qw              &\qw       &\qw                   &\qw              &\qw       &\qw                    &                       &\qw\\
\qw         &\qw\qwbundle{1}&\gate[wires=2]{K}&\qw       &\qw                   &\gate[wires=2]{K}&\qw       &\qw                   &\gate[wires=2]{K}&\qw       &\qw                    &\qw                    &\qw\\
\qw         &\qw\qwbundle{4}&                 &\qw       &\octrl{-2}\qwbundle{4}&                 &\qw       &\octrl{-3}\qwbundle{4}&                 &\qw       &\octrl{-3}\qwbundle{4} &\qw                    &\qw
\end{quantikz}     
\caption{The diagram for the circuit for an amplitude estimation with two output qubits. The gates $K$ are the parts on the left-hand side of the separation line of the circuit in figure~\ref{figureErrorMitigation}. Here the first and second qubits in the circuit control only the $G$ operation inside $K$. Note that the gates $K$ contain non-unitary reset operations.}
\label{phaserExample3qubits}
\end{figure}
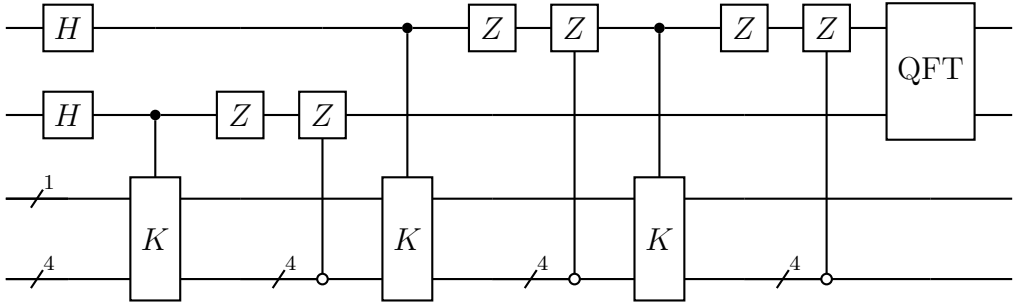

To see how our techniques perform in the presence of errors we use a simple error model with the error probability $p_{\rm error}$ for each application of controlled Grover operators. In this model, we insert an $X$ gate before the EP gate with probability $p_{\rm error}/2$ and before the Grover operator with probability $p_{\rm error}$. The $X$ gate acts on a qubit that is randomly chosen from the register of the Grover operator and the control qubit. The results of 1000 experiments with $p_{\rm error}=0.2$ for a QAE with 5 bit resolution look like follows:

\begin{figure}[H]
\centering
\includegraphics[width=1.0\textwidth]{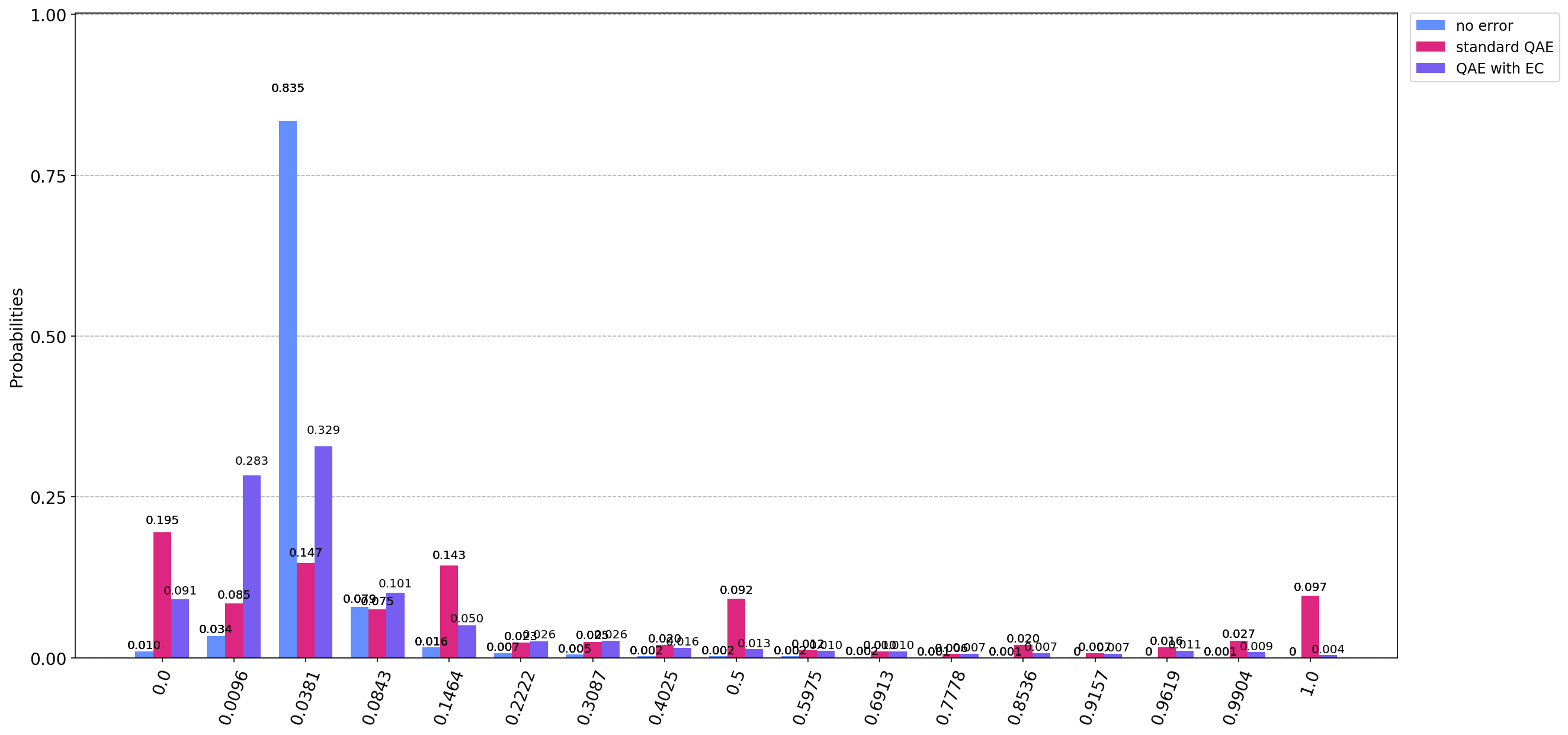}
\caption{The results of standard QAE and the error-corrected version of the parallel QAE for $p_{\rm error}=0.2$ and $5$ bit precision. For comparison, the error-free case is also included. We generated $1000$ random errors and sampled each with $100$ shots.}
\label{histo}
\end{figure}

We see that in the error free case, which we included for comparison, the probability is mostly in the bin corresponding to a probability of 3.8\% for the worst case event. We linearly interpolate the angles corresponding to the two most likely bins found by the QAE. In this case, this means a linear interpolation between the probability of 3.8\% and 7.9\%. The resulting probability is 4.1\%.\\

For the standard QAE with errors, the highest peak is in the 0-th bin, indicating a 0\% probability for the worst case event. The second highest peak is not in a neighboring bin, but two bins away. Interpolation does not seem to make sense here, especially given the fact that the third highest bin is 4 bins away, and there even are significant peaks at 50\% and 100\%. In summary, we see no way how to get a useful result from these measurements.\\

Now we consider the parallel QAE: The two highest peaks are in the second and third bin. An interpolation as above leads to a result of 2.2\% probability for the worst case scenario of our risk model.
We can improve our result further with an estimation of the error rate $p_{\rm error}$, because we know that errors shift the measured angle by the factor $(1 -p_{\rm error})$. In our case, we can build and measure a circuit which executes the Grover operator, followed by its inverse. The instances where we do not measure all qubits in the state $|0\rangle$ corresponds to an error. This procedure would yield $p_{\rm error}=0.2$, and the result would be shifted by a factor of $0.9\cdot 0.8=0.72$ since the EP-gate would contribute errors with $p_{\rm error}/2$.
If we correct with this factor, our final result is a 4.3\% probability.\\

A difference between the error free case and the error corrected case is the probability that is concentrated in the two bins with highest probability.
It is 90\% for the error free case and about 60\% for the parallel case. This means that we have to measure the circuit more often than in the error free case to get a good result.\\

In summary, the parallel setup gave a result of 4.3\% probability for the worst case scenario on simulated quantum hardware with errors. For comparison, the error free version gives 4.1\% probability, while the true result is 4.7\%. The standard QAE fails to produce a useful answer with this noise model.

\section{Conclusion and Outlook}
We discussed several methods for the parallelization of phase and amplitude estimation. The indirect parallelization is of particular interest because it reduces the gate depth of such quantum circuits significantly. It shortens the calculation time on hardware platforms that support the parallel execution of operations and also helps to reduce problems with limited coherence times.\\

The obvious difficulty to overcome for the parallelization of amplitude estimation is the preparation of single eigenstates of Grover operators. We demonstrated how this can be achieved for some cases in section~\ref{secEP}. After finishing the first version of this paper, 
we became aware of the earlier work~\cite{Knill}, where the parallel construction of section~\ref{secIndParallel} is described. In this work, a different construction of the eigenstates is given and this method complements ours, because the cases for which it works efficiently is different from ours.

We also analyzed the effect of errors on the results. It turns out that as long as errors happen outside of the Grover operators, the parallel circuits show a much better error resilience than serial implementations. We showed this with simulated circuits and derived the theory for this. The results are summarized in figure~\ref{plotLDcomparison3}. However, on real, noisy quantum hardware the errors should generally be expected to appear inside the Grover operators. In such cases, we find the performance of serial and parallel circuits to be quite similar. We explain the theoretical origin for this in section~\ref{errorsingrovers}. The consequences of errors inside Grover operators also mean that error correction methods as outlined in section~\ref{sectionErrorCorrection} will not work in such settings. What still works is the flagging of errors, which can be used to post-select results from correctly executed circuits.

\section*{Acknowledgements}
We thank Fred Jendrzejewski, Thomas Monz and Stefan Woerner for useful discussions.

\appendix

\section{Calculations for Operators on One Qubit}\label{sectionOneQubit}
Operators on one qubit form a special class regarding the error calculations, because the entire state space is spanned by the two Grover eigenvectors and there are no additional eigenstates with eigenvalues $+1$ or $-1$. Therefore, the error analysis is quite different compared to the systems with more than one qubit that are analyzed above.

\subsection{Standard Amplitude Estimation}
As pointed out in section~\ref{errorsin2d}, in the case of Grover operators on one qubit a bit or phase flip error changes the state of the Grover register from one eigenstate to the other. This implies that the phase that the operator kicks back changes its sign.\\

It is instructive to think about the accumulation of phase kickbacks as a random walk. For example, assume that the register collecting the kickbacks is initialized in the state 
$$
\frac{1}{\sqrt{2}}\left(|0\rangle + |1\rangle \right)
$$
and as we move through the series of Grover operators, each one kicks back $+2\theta$. As soon as an error occurs, the state flips to the other eigenvector and the kickback is $-2\theta$ for all following Grover operators until another error occurs and the state changes back again. This is the behavior of a random walk, where we flip a coin and we keep moving in the same direction when the coin comes up head and we change the direction when it comes up tails. This is known as an one dimensional persistent random walk~\cite{persRW1,persRW2}.\\

In terms of the random walk, we want to know how far from the starting point we expect to be after $N$ steps. This corresponds to the number of kickbacks we have accumulated after $N$ Grover operators. If our random walk was not persistent, we could have 
used the known formula
\begin{equation*}
d = \sqrt{\frac{2 N}{\pi}}
\end{equation*}
for the mean distance from the starting point, see \cite{rwdistance1, rwdistance2, rwdistance3}.\\

In our case, we can describe the persistent random walk as a standard random walk with a modified step size. Since the number of steps before a direction change is described by a geometric distribution, the expected number of steps before a change is $1/p$. We denote the number of these new steps as $\tilde{N}$. The role of the variance $\sqrt{N}$ in a standard random walk is taken over by $\sqrt{\tilde{N}}$, which is the variance of the binomial distribution of direction changes:
\begin{equation}
\sqrt{\tilde{N}} = \sqrt{p (1-p) N}
\end{equation}
If we multiply this with the expected size $1/p$ of the rescaled steps of the persistent random walk, then we find the distance $\tilde{d}$. This is what we expect 
the persistent random walk to be away from the starting point after $N$ steps. It is given by 
\begin{equation*}
\tilde{d} = \frac{1}{p} \sqrt{\frac{2 \tilde{N}}{\pi}} =  \sqrt{\frac{\left(\frac{1}{p}-1\right) 2 N}{\pi}}
\end{equation*}
Note that our approximation only makes sense as long as $1/p$ is smaller than $N$.\\
This shows that the behavior of QAE or low depth QAE for oracles with one qubit is better than the general case, because there is no bound like equation~\ref{Nbound}. Nevertheless, the number of kickbacks that the circuit accumulates does not scale like $N$, but is proportional to $\sqrt{N}$. Therefore, the asymptotic scaling is not better than that of classical Monte-Carlo methods up to constant factors.

\subsection{Parallel Amplitude Estimation}
In the parallel setup, the situation is similar to that of Grover operators on more than one qubit. The only difference is that in the case of an error the kickback angle is  $-2\theta$ instead of the correct kickback angle $2 \theta$~\footnote{Note that in the higher dimensional case, the kickback simply was an angle of $0$ or $\pi$ corresponding to the phase $+1$ or $-1$.}.
The parallel setup can be described by a standard random walk with drift. A successful execution of the Grover operator is a step to the right, and a faulty operator execution corresponds to a step to the left. The positions are distributed according to a binomial distribution. In terms of the expected number $k$ of kickbacks of the angle $2\theta$, we can write the expectation after $N$ steps as
$$
\mathbb{E}(2\theta k) = 2 \theta (1-2p) N
$$
This shows that the effect of an error is doubled compared to the case with more than one qubit, see equation~\ref{parallelkicks}. Qualitatively, the behavior of parallel QAE is the same for Grover operators on one qubit and for Grover operators on more than one qubit.

\bibliographystyle{ieeetr}
\bibliography{refs2}

\end{document}